\documentclass[preprint,prd,nofootinbib,tightenlines,amsmath,amssymbol]{revtex4}
\usepackage{graphicx}
\usepackage{color}

\begin{document}
\baselineskip=15pt \parskip=5pt

\vspace*{3em}

\preprint{}

\title{Probing New Physics in Charm Couplings with FCNC}

\author{Xiao-Gang He}
\email{hexg@phys.ntu.edu.tw}
\affiliation{Department of Physics and Center for Theoretical Sciences, \\
National Taiwan University, Taipei 106, Taiwan}

\author{Jusak Tandean}
\email{jtandean@yahoo.com}
\affiliation{Department of Physics and Center for Theoretical Sciences, \\
National Taiwan University, Taipei 106, Taiwan}

\author{G. Valencia}
\email{valencia@iastate.edu}
\affiliation{Department of Physics and Astronomy, Iowa State University, Ames, Iowa 50011, USA}

\date{\today $\vphantom{\bigg|_{\bigg|}^|}$}

\begin{abstract}

Low-energy experiments involving kaon, $B$-meson, $D$-meson, and hyperon flavor-changing neutral
transitions have confirmed the loop-induced flavor-changing neutral current (FCNC) picture of
the standard model~(SM).
The continuing study of these processes is essential to further refine this picture and ultimately
understand the flavor dynamics.
In this paper we consider deviations from the SM in the charm sector and their effect on FCNC
processes.  Specifically, we parametrize new physics in terms of left- and right-handed anomalous
couplings of the $W$ boson to the charm quark.
We present a comprehensive study of existing constraints and point out those measurements that
are most sensitive to new physics of this~type.

\end{abstract}

\pacs{PACS numbers: }

\maketitle

\section{Introduction}

One of the outstanding problems for high-energy physics remains the understanding of the dynamics
of flavor.  Existing experimental results on kaon, $D$-meson, $B$-meson, and hyperon decays,
as well as neutral-meson mixing, are all consistent with the loop-induced nature of flavor-changing
neutral currents (FCNC's) in the standard model~(SM) and also with the unitarity of
the Cabibbo-Kobayashi-Maskawa (CKM) matrix with three generations.  The continuing study of these
processes with increased precision will play a crucial role in the search for physics beyond the~SM.

In many types of new physics, the new particles are heavier than their SM counterparts and
their effects can be described by an effective low-energy theory.
A complete set of operators of dimension six describing deviations from the SM has been
presented in Ref.~\cite{Buchmuller:1985jz}.
A less ambitious program is to study only those operators that appear when the new physics
effectively modifies the SM couplings between gauge bosons and certain fermions~\cite{Peccei:1989kr}.
The case of anomalous top-quark couplings has been treated before  in
the literature~\cite{Fujikawa:1993zu,anomtqc}, and it was found that they are most tightly
constrained by the \,$b\to s\gamma$\, decay.  Interestingly, this mode does not place severe
constraints on anomalous charm-quark couplings due to the relative smallness of the charm mass.

In this paper we focus on new physics affecting primarily the charged weak currents involving
the charm quark.
Including the SM term, the effective Lagrangian in the unitary gauge for a~general parametrization
of anomalous interactions of the $W$ boson with an up-type quark $U_k^{}$ and a~down-type quark
$D_l^{}$ can be written as
\begin{eqnarray} \label{ludw}
{\cal L}_{UDW}^{} \,\,=\,\,  -\frac{g}{\sqrt2}\, V_{kl}^{}\,\bar U_k^{}\gamma^\mu \bigl[
\bigl(1+\kappa_{kl}^{\rm L}\bigr)P_{\rm L}^{}
\,+\, \kappa_{kl}^{\rm R}\,P_{\rm R}^{} \bigr] D_l^{}\, W_\mu^+
\,\,+\,\, {\rm H.c.} \,\,,
\end{eqnarray}
where $g$ is the weak coupling constant, we have normalized the anomalous couplings
$\kappa^{\rm L,R}_{kl}$ relative to the usual CKM-matrix elements $V_{kl}^{}$, and
\,$P_{\rm L,R}^{}=\frac{1}{2}(1\mp\gamma_5^{})$.\,
In general, $\kappa^{\rm L,R}_{kl}$ are complex and, as such, provide new sources of $CP$
violation.  In Appendix~\ref{Keff} we discuss the general parametrization of the quark-mixing
matrix underlying Eq.~(\ref{ludw}), paying particular attention to the number of independent
parameters that are allowed.

In addition to affecting weak decays though tree-level interactions, the new couplings in
Eq.~(\ref{ludw}) modify effective flavor-changing and -conserving couplings at one loop.
In this work, we evaluate several one-loop transitions induced by the new couplings via
magnetic-dipole, penguin, and box diagrams. We include the operators generated this way in
our phenomenological analysis.

We present a comprehensive picture of existing constraints which shows that deviations from SM
couplings at the percent level are still possible, particularly for right-handed interactions.
Our study will also serve as a guide as to which future measurements provide the most sensitive
tests for new physics that can be parametrized with anomalous $W$-boson couplings to the charm quark.
For the $CP$-violating (imaginary) parts of the couplings, the electric dipole moment of the neutron
and the hyperon asymmetry $A_{\Xi\Lambda}^{}$ are the most promising channels to probe for
right-handed couplings, whereas more precise measurements of $\sin(2\beta)$ and $\sin(2\beta_s)$
are the most promising probes for left-handed couplings.
Constraints on the real parts of the right-handed couplings can be further improved with better
measurements of semileptonic $B$ and $D$ decays.

\section{One-loop processes\label{1loop}}

In this section, we collect the main formulas for the loop-induced processes we are considering.
All our calculations are performed in the unitary gauge, and the results have been compared to
existing ones, where available. We have summarized our loop calculations in Appendix~\ref{unitary}.

\subsection{Dipole penguin operators}

Of particular interest are the electromagnetic and chromomagnetic dipole operators, which can
give rise to potentially large corrections to SM processes~\cite{Kagan:1994qg,Buras:1999da} and
be expressed as
\begin{eqnarray}
Q_\gamma^\pm  \,\,=\,\, \frac{e}{16\pi^2} \bigl( \bar d'\sigma_{\mu\nu}^{}P_{\rm R}^{}d
\pm \bar d'\sigma_{\mu\nu}^{}P_{\rm L}^{}d\bigr) F^{\mu\nu}   \,\,,
\end{eqnarray}
\begin{eqnarray}
Q_g^\pm  \,\,=\,\,
\frac{g_{\rm s}^{}}{16\pi^2} \bigl(\bar d'\sigma_{\mu\nu}^{}t_a^{}P_{\rm R}^{}d
\pm \bar d'\sigma_{\mu\nu}^{}t_a^{}P_{\rm L}^{}d\bigr) G_a^{\mu\nu} \,\,,
\end{eqnarray}
where  $d$ and \,$d'\neq d$\, are down-type quarks, $F^{\mu\nu}$ and  $G_a^{\mu\nu}$ are the usual
photon and gluon field-strength tensors, respectively, with $e$ and $g_{\rm s}^{}$ being
their coupling constants, and  $\,{\rm Tr}(t_a t_b)=\frac{1}{2}\delta_{ab}.\,$
The main effect of these operators is generated at one loop with the $W$ boson coupling to
the left-handed current at one vertex and to the right-handed current at the other vertex.
It corresponds to the terms linear in the $\kappa^{\rm R}$'s in Eqs.~(\ref{Hdd'f})
and~(\ref{Hdd'g}), leading to the effective Hamiltonian
\begin{eqnarray} \label{Hg}
{\cal H}_{\gamma,g}^{} \,\,=\,\, C_\gamma^+ Q_\gamma^+ + C_\gamma^- Q_\gamma^- +
C_g^+ Q_g^+ + C_g^- Q_g^-  \,\,+\,\,  {\rm H.c.}   \,\,,
\end{eqnarray}
where at the $W$-mass scale
\begin{eqnarray} \label{Cg}
\begin{array}{c}   \displaystyle
C_\gamma^\pm(m_W^{}) \,\,=\,\, -\sqrt2\,G_{\rm F}^{} \sum_{q=u,c,t} V_{qd'}^{*} V_{qd}^{}\,
\bigl( \kappa_{qd}^{\rm R}\pm\kappa_{qd'}^{\rm R*} \bigr) \, m_q^{}\,F_0^{}(x_q^{})   \,\,,
\vspace{2ex} \\   \displaystyle
C_g^\pm(m_W^{})  \,\,=\,\, -\sqrt2\,G_{\rm F}^{} \sum_{q=u,c,t} V_{qd'}^{*} V_{qd}^{}\,
\bigl( \kappa_{qd}^{\rm R}\pm\kappa_{qd'}^{\rm R*} \bigr) \, m_q^{}\, G_0^{}(x_q^{}) \,\,,
\end{array}
\end{eqnarray}
with  $G_{\rm F}^{}$ being the Fermi constant, \,$x_q^{}=\bar m_q^2\bigl(m_q^{}\bigr)/m_W^2$,\,
and $F_0^{}$ and $G_0^{}$ given in Eqs.~(\ref{f}) and~(\ref{g}).
The contributions of these operators are potentially enlarged relative to
the corresponding ones in the SM due to the enhancement factors of $m_c^{}/m_s^{}$ and
$m_t^{}/m_b^{}$ in \,$s\to d$\, and \,$b\to(d,s)$\, transitions, respectively, and also due
to $F_0^{}(x_q^{})$ and $G_0^{}(x_q^{})$ being larger than their SM counterparts.
For the case of anomalous $tbW$ couplings, the formulas above agree with those found in
the literature~\cite{Fujikawa:1993zu}.

\subsection{Electric dipole moments}

The flavor-conserving counterparts of $Q_{\gamma,g}$ above contribute to the electric and color
dipole-moments of the $d$ and $s$ quarks.  Based on Eqs.~(\ref{Hg}) and~(\ref{Cg}), one can write
the effective Hamiltonian for such contributions to the dipole moments of the $d$ quark as
\begin{eqnarray} \label{Hd}
{\cal H}_d^{\rm dm} \,\,=\,\,
\mbox{$\frac{i}{2}$}\, d_d^{\rm edm}\,\bar d\sigma_{\mu\nu}^{}\gamma_5^{}d\, F^{\mu\nu}
\,+\, \mbox{$\frac{i}{2}$}\,
d_d^{\rm cdm}\, \bar d\sigma_{\mu\nu}^{}t_a^{}\gamma_5^{}d\, G_a^{\mu\nu} \,\,,
\end{eqnarray}
where in our case
\begin{eqnarray} \label{dedm}
\begin{array}{c}   \displaystyle
d_d^{\rm edm}(m_W^{}) \,\,=\,\,
\frac{-e G_{\rm F}^{}}{2\sqrt2\,\pi^2} \sum_{q=u,c,t} \bigl|V_{qd}^{}\bigr|^2\,
{\rm Im}\,\kappa_{qd}^{\rm R}\, m_q^{}\, F_0^{}(x_q^{}) \,\,,
\vspace{2ex} \\   \displaystyle
d_d^{\rm cdm}(m_W^{}) \,\,=\,\,
\frac{-g_{\rm s}^{} G_{\rm F}^{}}{2\sqrt2\,\pi^2} \sum_{q=u,c,t} \bigl|V_{qd}^{}\bigr|^2\,
{\rm Im}\,\kappa_{qd}^{\rm R}\, m_q^{}\, G_0^{}(x_q^{})
\end{array}
\end{eqnarray}
at the $m_W^{}$ scale.
These expressions agree with those derived from quark-$W$ loop diagrams in a~left-right
model~\cite{Ecker:1983dj}, after appropriate changes are made.
The corresponding quantities for the $s$ quark are similar in form.
Since the dipole moments $d_{d,s}^{}$ contribute to the electric dipole moment of the
neutron~\cite{He:1989xj,Ginges:2003qt,Dib:2006hk}, it can be used to place constraints on
${\rm Im}\,\kappa_{qd,qs}^{\rm R}$.

\subsection{\boldmath Operators generated by $Z$-penguin, $\gamma$-penguin, and box diagrams}

The anomalous quark-$W$ couplings also generate flavor-changing neutral-current interactions
via $Z$-penguin, $\gamma$-penguin, and box diagrams.
They will therefore affect other loop-generated processes, such as \,$K\to\pi\nu\bar\nu$,\,
\,$K_L\to\ell^+\ell^-$,\, and neutral-meson mixing.

The effective theory with anomalous couplings is not renormalizable, and this results in
divergent contributions to some of the processes we consider. These divergences are understood in
the context of effective field theories as contributions to the coefficients of higher-dimension
operators.  These operators then enter the calculation as additional `anomalous couplings',
introducing new parameters to be extracted from experiment.
For our numerical analysis, we will limit ourselves to the anomalous couplings of Eq.~(\ref{ludw}),
ignoring the higher-dimension operators.
In so doing, we trade the possibility of obtaining precise predictions in specific models for
order-of-magnitude estimates of the effects of new physics parametrized in a model-independent way.
We will rely on the common procedure~\cite{Georgi:1994qn} of using dimensional regularization,
dropping the resulting pole in four dimensions, and identifying the renormalization scale~$\mu$
with the scale of the new physics underlying the effective theory. Our results will thus contain
a logarithmic term of the form  $\ln\bigl(\mu/m_W^{}\bigr)$ in which
we set \,$\mu=\Lambda=1$\,TeV\, for definiteness.
In addition to the logarithmic term representing the new-physics contribution, we have also kept
in our estimates those finite terms that correspond to contributions from SM quarks in the
loops.\footnote{Explicit examples of the type of divergence cancelation resulting in a~logarithmic
term as described above can be found in Ref.~\cite{He:2004it}, where the new-physics scale is given
by the masses of non-SM Higgs-bosons.}

We consider the contributions of the anomalous couplings to \,$d\bar d'\to\nu\bar\nu$,\,
\,$d\bar d'\to\ell^+\ell^-$,\,  and  \,$d\bar d'\to\bar d d'$,\, relegating the main results
of the calculation to Appendix~\ref{unitary}.
It follows that the effective Hamiltonians generated by the anomalous charm couplings are at
the $m_W^{}$ scale
\begin{eqnarray} \label{hnunu}
{\cal H}_{d\bar d'\to\nu\bar\nu}^{\kappa} &=&
\frac{\alpha\, G_{\rm F}^{}\, \lambda_c^{}\,\bigl(\kappa_{cd}^{\rm L}+\kappa_{cd'}^{\rm L*}\bigr)}
     {\sqrt8\, \pi\, \sin^2\theta_{\rm W}^{}}
\biggl( -3\, \ln\frac{\Lambda}{m_W^{}} + 4 X_0^{}\bigl(x_c^{}\bigr) \biggr)
\bar d'\gamma^\sigma P_{\rm L}^{}d\, \bar\nu\gamma_\sigma^{}P_{\rm L}^{}\nu
\nonumber \\ && \!\!\! +\,\,
\frac{\alpha\, G_{\rm F}^{}\, \lambda_c^{}\,\kappa_{cd}^{\rm R}\kappa_{cd'}^{\rm R*}}
     {\sqrt8\, \pi\, \sin^2\theta_{\rm W}^{}}
\biggl[ \bigl(4 x_c^{}-3\bigr)\, \ln\frac{\Lambda}{m_W^{}} + \tilde X\bigl(x_c^{}\bigr) \biggr]
\bar d'\gamma^\sigma P_{\rm R}^{}d\, \bar\nu\gamma_\sigma^{}P_{\rm L}^{}\nu \,\,,
\end{eqnarray}
\begin{eqnarray} \label{hll}
{\cal H}_{d\bar d'\to\ell^+\ell^-}^{\kappa} &=&
\frac{\alpha\, G_{\rm F}^{}\, \lambda_c^{}\,\bigl(\kappa_{cd}^{\rm L}+\kappa_{cd'}^{\rm L*}\bigr)}
     {\sqrt8\, \pi}
\Biggl[ \biggl( 3\, \ln\frac{\Lambda}{m_W^{}} - 4 Y_0^{}\bigl(x_c^{}\bigr) \biggr)
\frac{\bar d'\gamma^\sigma P_{\rm L}^{}d\,\bar\ell\gamma_\sigma^{} P_{\rm L}^{}\ell}
     {\sin^2\theta_{\rm W}^{}}
\nonumber \\ && \hspace*{21ex} +\,\,
\biggl(-\frac{16}{3}\, \ln\frac{\Lambda}{m_W^{}} + 8 Z_0^{}\bigl(x_c^{}\bigr) \biggr)
\bar d'\gamma^\sigma P_{\rm L}^{}d\, \bar\ell\gamma_\sigma^{}\ell \Biggr]
\nonumber \\ && \!\!\! \! +\,\,
\frac{\alpha\, G_{\rm F}^{}\, \lambda_c^{}\,\kappa_{cd}^{\rm R}\kappa_{cd'}^{\rm R*}}{\sqrt8\,\pi}
\Biggl\{\biggl[\bigl(3-4 x_c^{}\bigr)\,\ln\frac{\Lambda}{m_W^{}}+\tilde Y\bigl(x_c^{}\bigr)\biggr]
\frac{\bar d'\gamma^\sigma P_{\rm R}^{}d\,\bar\ell\gamma_\sigma^{} P_{\rm L}^{}\ell}
     {\sin^2\theta_{\rm W}^{}}
\nonumber \\ && \hspace*{17ex} +\,\,
\biggl[\biggl(8 x_c^{}-\frac{16}{3}\biggr)\,\ln\frac{\Lambda}{m_W^{}}+\tilde Z\bigl(x_c^{}\bigr)\biggr]
\bar d'\gamma^\sigma P_{\rm R}^{}d\, \bar\ell\gamma_\sigma^{}\ell \Biggr\} \,\,,
\end{eqnarray}
\begin{eqnarray} \label{hbox}
{\cal H}_{d\bar d'\to\bar d d'}^\kappa &=&
\frac{G_{\rm F}^2\,m_W^2}{8\pi^2}\, \lambda_c^{}\,\bigl(\kappa_{cd}^{\rm L}+\kappa_{cd'}^{\rm L*}\bigr)
\left( -\lambda_t^{}\, x_t^{}\,\ln\frac{\Lambda^2}{m_W^2} \,-\,
\sum_q\lambda_q^{}\, {\cal B}_1^{}\bigl(x_q^{},x_c^{}\bigr)
\right) \bar d'\gamma^\alpha P_{\rm L}^{}d\,\bar d'\gamma_\alpha^{}P_{\rm L}^{}d
\nonumber \\ && \!\!\!\! +\,\,
\frac{G_{\rm F}^2\,m_W^2}{4\pi^2}\, \lambda_c^{}\,\kappa_{cd}^{\rm R}\kappa_{cd'}^{\rm R*}
\left( -\lambda_t^{}\,x_t^{}\,\ln\frac{\Lambda^2}{m_W^2} \,-\,
\sum_q\lambda_q^{}\, {\cal B}_2^{}\bigl(x_q^{},x_c^{}\bigr)
\right) \bar d'\gamma^\alpha P_{\rm L}^{}d\,\bar d'\gamma_\alpha^{}P_{\rm R}^{}d
\nonumber \\ && \!\!\!\! +\,\,
\frac{G_{\rm F}^2\,m_W^2}{4\pi^2}\, \lambda_c^2\, x_c^{}\,
\biggl( -\ln\frac{\Lambda^2}{m_W^2} \,-\, {\cal B}_3^{}\bigl(x_c^{},x_c^{}\bigr) \biggr)
\Bigl[ \bigl(\kappa_{cd}^{\rm R}\bigr)^2\, \bar d'P_{\rm R}^{}d\,\bar d'P_{\rm R}^{}d +
\bigl(\kappa_{cd'}^{\rm R*}\bigr)^2\,\bar d'P_{\rm L}^{}d\,\bar d'P_{\rm L}^{}d \Bigr]
\nonumber \\
\end{eqnarray}
where \,$d'\neq d$,\, we have kept terms linear in $\kappa^{\rm L}$ and quadratic in $\kappa^{\rm R}$,
\,$\lambda_q^{}=V_{qd'}^*V_{qd}^{}$,\,  and  $\theta_{\rm W}^{}$ is the Weinberg angle.
The functions $X_0^{}$, $Y_0^{}$, $Z_0^{}$, $\tilde X$, $\tilde Y$, $\tilde Z$, and
${\cal B}_{1,2,3}$ can be found in Appendix~\ref{unitary}.

\section{Tree-level constraints}

From now on, we focus on the anomalous charm couplings  $\kappa_{cd,cs,cb}^{\rm L,R}$, neglecting
the corresponding $u$ and $t$ anomalous couplings.
To obtain constraints on the couplings, we begin by exploring their tree-level contributions to
three different sets of processes, \,$D_{(s)}\to\ell\nu$,\,  exclusive and inclusive
\,$b\to c\ell^-\bar\nu$\, transitions,  and  mixing-induced $CP$ violation in  \,$B\to J/\psi K$\,
and \,$B\to\eta_c^{}K$,\,  where the couplings may play some interesting roles.

\subsection{\boldmath$D,D_s\to\ell\nu$}

From the Lagrangian in Eq.~(\ref{ludw}), at tree level one derives the decay rate
\begin{eqnarray}
\Gamma(D\to\ell\nu) \,\,=\,\,
\frac{G_F^2\, f_D^2\, m_\ell^2\, m_D^{}}{8\pi}
\Biggl(1-\frac{m_\ell^2}{m_D^2}\Biggr)^{\!\!2}\,
\bigl|V_{cd}^{}\bigl(1+\kappa_{cd}^{\rm L}-\kappa_{cd}^{\rm R}\bigr)\bigr|^2 \,\,,
\end{eqnarray}
where the decay constant $f_D^{}$ is defined by
\,$\langle0|\bar d\gamma^\mu c|D(p)\rangle=i f_D^{}p^\mu$.\,
Changing $V_{cd}^{}$, $\kappa^{\rm L,R}_{cd}$, $m_D^{}$, and $f_D^{}$ to
$V_{cs}^{}$, $\kappa^{\rm L,R}_{cs}$, $m_{D_s}$ and $f_{D_s}$, respectively, one obtains the decay
width  $\Gamma(D_s^{}\to\ell\nu)$.

Recent measurements of \,$D,D_s\to\ell\nu$\, yield~\cite{pdg,Stone:2008gw}
\begin{eqnarray}
f_D^{\rm exp} \,\,=\,\, (205.8 \pm 8.9){\rm\,MeV} \,\,, \hspace{5ex}
f_{D_s^{}}^{\rm exp} \,\,=\,\, (261.2 \pm 6.9){\rm\,MeV}  \,\,,
\end{eqnarray}
whereas SM calculations give~\cite{Dobrescu:2008er,Narison:2008bc}
\begin{eqnarray}
f_D^{\rm th} \,\,=\,\, (202\pm8){\rm\,MeV} \,\,, \hspace{5ex}
f_{D_s^{}}^{\rm th} \,\,=\,\, (240\pm7){\rm\,MeV}  \,\,.
\end{eqnarray}
Evidently, for \,$D\to\ell\nu$\, the data agree with theoretical predictions well, but for
\,$D_s\to\ell\nu$\, there is deviation at the 2-sigma level.
It has been argued that this deviation may be due to physics beyond the SM~\cite{Dobrescu:2008er},
but it is too early to conclude that new physics is needed.

Nevertheless, one can turn the argument around to constrain new physics by assuming that
the discrepancy between the calculated and measured values of the decay constants arose from
the anomalous couplings, as the  $\Gamma\bigl(D_{(s)}\to\ell\nu\bigr)$  formulas would imply.
Using the experimental and theoretical numbers above, one can then extract
\begin{eqnarray} \label{recd}
\bigl|{\rm Re}\bigl(\kappa_{cd}^{\rm L}-\kappa_{cd}^{\rm R}\bigr)\bigr| \,\,\le\,\, 0.04 \,\,,
\end{eqnarray}
\begin{eqnarray} \label{recs}
0 \,\,\le\,\, {\rm Re}\bigl(\kappa_{cs}^{\rm L}-\kappa_{cs}^{\rm R}\bigr) \,\,\le\,\, 0.1 \,\,.
\end{eqnarray}

\subsection{\boldmath Semileptonic $B$ decay and extraction of $V_{cb}$}

The interaction in Eq.~(\ref{ludw}) will also affect the extraction of $V_{cb}$ from semileptonic
$B$ decay.  At the quark level, the effect of the new couplings is to scale the hadronic vector
and axial-vector currents by the factors \,$1+\kappa^{\rm L}_{cb}\pm\kappa^{\rm R}_{bc}$,\,
respectively.  This has the following implications.

First, the semileptonic exclusive decay  \,$\bar B\to D e\bar\nu_e^{}$\, is sensitive only to
the vector form-factor, and thus the differential (and total) decay rate simply gets multiplied
by   \,$\bigl|1+\kappa^{\rm L}_{cb}+\kappa^{\rm R}_{bc}\bigr|{}^2$.\,
To linear order in the $\kappa$'s, this means that what is measured in this mode is
\begin{eqnarray} \label{vc}
V_{cb}^{\rm eff} \,\,=\,\,
V_{cb}^{}\,\bigl(1+{\rm Re}\,\kappa^{\rm L}_{cb}+{\rm Re}\,\kappa^{\rm R}_{cb}\bigr) \,\,=\,\,
(39.4 \pm 4.4) \times 10^{-3} \,\,.
\end{eqnarray}
The number above and the other ones below for $V_{cb}^{\rm eff}$ are quoted from
Ref.~\cite{Kowalewski:2008zz}, and their errors result from adding the experimental and
theoretical uncertainties given therein in quadrature.

Second, the semileptonic exclusive decay \,$\bar B\to D^* e\bar\nu_e^{}$\, is sensitive to both
the vector and axial-vector currents.
In the heavy-quark limit, \,$w=v\cdot v'=1$\, (where $v$ and $v'$ are the four-velocities of
the $B$ and $D^*$, respectively), only the axial-vector current survives~\cite{Isgur:1989vq},
and so the decay rate in this limit would simply get multiplied by
\,$\bigl|1+\kappa^{\rm L}_{cb}-\kappa^{\rm R}_{bc}\bigr|{}^2$.\,
One can do better than this by considering the form factors in the heavy-quark effective theory (HQET)
where they either vanish or can be written in terms of the Isgur-Wise function  $\xi(w)$ with
the normalization \,$\xi(1)=1$\,~\cite{Isgur:1989vq}.
Treating the form factors as constants throughout the kinematically allowed range \,$1\le w\le1.5$,\,
one then finds to linear order in $\kappa$
\begin{eqnarray} \label{avc}
V_{cb}^{\rm eff} \,\,=\,\,
V_{cb}^{}\,\bigl(1+{\rm Re}\,\kappa^{\rm L}_{cb}-0.93{\rm Re}\,\kappa^{\rm R}_{cb}\bigr)
\,\,=\,\, (38.6 \pm 1.4) \times 10^{-3} \,\,.
\end{eqnarray}

Third, the semileptonic inclusive decay rate can be easily calculated to be
\begin{eqnarray}
\Gamma\bigl(b\to c e^-\bar\nu_e^{}\bigr) \,\,=\,\,
\frac{G_{\rm F}^2\,m_b^5}{192\pi^3}|V_{cb}^{}|^2 \Bigl\{
F(r)\,\Bigl(\bigl|1+\kappa^{\rm L}_{cb}\bigr|^2+\bigl|\kappa^{\rm R}_{cb}\bigr|^2\Bigr)
+ 2G(r)\,{\rm Re}\bigl[\bigl(1+\kappa^{\rm L}_{cb}\bigr)\kappa^{\rm R*}_{cb}\bigr]\Bigr\} \,\,,
\end{eqnarray}
where \,$r=m_c^{}/m_b^{}\simeq0.3$,\,
\begin{eqnarray}
F(r) &=& 1-8r^2+8r^6-r^8-24r^4\,\ln r \,\,, \nonumber \\
G(r) &=& -8r\,\bigl[1+9r^2-9r^4-r^6+12r^2\,\bigl(1+r^2\bigr)\, \ln r\bigr] \,\,.
\end{eqnarray}
It follows that to linear order in $\kappa$
\begin{eqnarray} \label{incl}
V_{cb}^{\rm eff} \,\,=\,\,
V_{cb}^{}\,\bigl(1+{\rm Re}\,\kappa^{\rm L}_{cb}-1.5\,{\rm Re}\, \kappa^{\rm R}_{cb}\bigr)
\,\,=\,\, (41.6 \pm 0.6) \times 10^{-3} \,\,.
\end{eqnarray}

From these results it is evident that it is not possible to extract a bound on
$\kappa^{\rm L}_{cb}$ (as long as quadratic effects are ignored), but we can extract bounds
on  $\kappa^{\rm R}_{cb}$.
For example, we can do a~two-parameter fit to Eqs.~(\ref{vc}), (\ref{avc}), and~(\ref{incl})
to find a $\chi^2$ minimum for
\begin{eqnarray} \label{fit}
V_{cb}^{}\,\bigl(1+{\rm Re}\,\kappa^{\rm L}_{cb}\bigr) \,\,=\,\, 0.038 \,\,, \hspace{5ex}
{\rm Re}\,\kappa^{\rm R}_{cb} \,\,=\,\, -0.057 \,\,,
\end{eqnarray}
with a corresponding 68\%-C.L. interval (1-$\sigma$ error)
\begin{eqnarray} \label{recb}
-0.13 \,\,\le\,\, {\rm Re}\,\kappa^{\rm R}_{cb} \,\,\le\,\, 0 \,\,.
\end{eqnarray}

\subsection{\boldmath$CP$ violation in  $B\to J/\psi K$ and $B\to\eta_c^{}K$}

One of the decay modes expected to provide a clean determination of the unitarity-triangle
parameter $\beta$ from the measurement of time-dependent $CP$ violation is \,$B\to\eta_c^{}K$,\,
just like \,$B\to J/\psi K$.\,
The SM predicts the same $\sin(2\beta)$ for the two processes, whereas the current data
for its effective values are~\cite{hfag}
\begin{eqnarray} \label{sin2b}
\sin\bigl(2\beta_{\psi K}^{\rm eff}\bigr) \,\,=\,\, 0.657 \pm 0.025 \,\,, \hspace{5ex}
\sin\bigl(2\beta_{\eta_c^{}K}^{\rm eff}\bigr) \,\,=\,\, 0.93 \pm 0.17 \,\,,
\end{eqnarray}
which disagree with each other at the 1.5-sigma level. Once again we can use the difference
between the two measurements to constrain the new physics parametrized by the anomalous couplings.
Since $\sin\bigl(2\beta^{\rm eff}\bigr)$ measures the difference between the phase of
the $B$-mixing matrix element and the phase of the ratio of amplitudes for the $B$ decay and its
antiparticle decay~\cite{Nir:1992wi}, then the discrepancy in $\beta^{\rm eff}$ between the two
modes must arise from a difference between the phases of their amplitude ratios.

The effective Hamiltonian for the \,$b\to sc\bar c$\, transition including the contribution of
anomalous couplings can be written as
\begin{eqnarray}
{\cal H}_{b\to sc\bar c}^{}  &=&
\frac{4G_{\rm F}^{}}{\sqrt2}\, V_{cs}^*V_{cb}^{}\, \Bigl(
C_1^{}\, \bar c\gamma^\mu P_{\rm L}^{}c\,\bar s\gamma_\mu^{}P_{\rm L}^{}b
+ C_2^{}\, \bar s\gamma^\mu P_{\rm L}^{}c\,\bar c\gamma_\mu^{}P_{\rm L}^{}b
\nonumber \\ && \hspace*{12ex} +\,\, \vphantom{|_{\big|}}
C_1^{\rm LR}\, \bar s_m^{}\gamma^\mu P_{\rm L}^{}c_n^{}\,\bar c_n^{}\gamma_\mu^{}P_{\rm R}^{}b_m^{}
+ C_2^{\rm LR}\, \bar s\gamma^\mu P_{\rm L}^{}c\,\bar c\gamma_\mu^{}P_{\rm R}^{}b
\nonumber \\ && \hspace*{12ex} +\,\,
C_1^{\rm RL}\, \bar s_m^{}\gamma^\mu P_{\rm R}^{}c_n^{}\,\bar c_n^{}\gamma_\mu^{}P_{\rm L}^{}b_m^{}
+ C_2^{\rm RL}\, \bar s\gamma^\mu P_{\rm R}^{}c\,\bar c\gamma_\mu^{}P_{\rm L}^{}b \Bigr) \,\,,
\end{eqnarray}
where $C_{1,2}^{\rm(LR,RL)}$ are the Wilson coefficients, $m$ and $n$ are color indices,
and we have neglected penguin operators.
To linear order in $\kappa$, the Wilson coefficients at the $m_W^{}$ scale are
\,$C_2^{}(m_W^{})=1+\kappa_{cs}^{\rm L*}+\kappa_{cb}^{\rm L}$,\,
\,$C_2^{\rm LR}(m_W^{})=\kappa_{cb}^{\rm R}$,\, and  \,$C_2^{\rm RL}(m_W^{})=\kappa_{cs}^{\rm R*}$.\,
These can be evolved down to a renormalization scale \,$\mu\sim m_b^{}$\, to
become~\cite{Buchalla:1995vs,Cho:1993zb}
\begin{eqnarray}
\begin{array}{c}   \displaystyle
C_1^{}(\mu) \,\,=\,\,
\mbox{$\frac{1}{2}$}\Bigl(\eta^{6/23}-\eta^{-12/23}\Bigr)C_2^{}(m_W^{})  \,\,, \hspace{5ex}
C_2^{}(\mu) \,\,=\,\, \mbox{$\frac{1}{2}$}\Bigl(\eta^{6/23}+\eta^{-12/23}\Bigr)C_2^{}(m_W^{})  \,\,,
\vspace{1ex} \\   \displaystyle
C_1^{\rm LR,RL}(\mu) \,\,=\,\,
\mbox{$\frac{1}{3}$}\Bigl(\eta^{-24/23}-\eta^{3/23}\Bigr)C_2^{\rm LR,RL}(m_W^{}) \,\,, \hspace{5ex}
C_2^{\rm LR,RL}(\mu) \,\,=\,\, \eta^{3/23}\, C_2^{\rm LR,RL}(m_W^{})
\end{array}
\end{eqnarray}
at leading order in QCD, where  \,$\eta=\alpha_{\rm s}^{}(m_W^{})/\alpha_{\rm s}^{}(\mu)$,

To determine the amplitudes for  \,$\bar B\to J/\psi\bar K,\eta_c^{}\bar K$,\,  we adopt the naive
factorization approximation.   The relevant matrix elements and parameter values are collected in
Appendix~\ref{me}.  It follows that
\begin{eqnarray} \label{amp}
\begin{array}{c}   \displaystyle
{\cal M}\bigl(\bar B^0\to\psi\bar K^0\bigr) \,\,=\,\,
\sqrt2\, G_{\rm F}^{}\,V_{cs}^*V_{cb}^{}\,
\bigl(1+\kappa_{cs}^{\rm L*}+\kappa_{cb}^{\rm L}\bigr)\, a_1^{}\, f_\psi^{} m_\psi^{}\, F_1^{BK}\,
\varepsilon_\psi\!\cdot\!p_K^{} \,\,,
\vspace{2ex} \\   \displaystyle
{\cal M}\bigl(\bar B^0\to\eta_c^{}\bar K^0\bigr) \,\,=\,\,
\frac{i G_{\rm F}^{}}{\sqrt2}V_{cs}^*V_{cb}^{} \left[
\bigl(1+\kappa_{cs}^{\rm L*}+\kappa_{cb}^{\rm L}\bigr) a_1^{} +
\frac{\bigl(a_1^{\rm LR}-a_1^{\rm RL}\bigr)m_{\eta_c^{}}^2}{m_c^{}\bigl(m_b^{}-m_s^{}\bigr)}
\right] \bigl(m_B^2-m_K^2\bigr) f_{\eta_c^{}}^{} F_0^{BK} \,\,,
\end{array}
\end{eqnarray}
where
\begin{eqnarray} \label{a1}
a_1^{} \,\,=\,\, C_1^{} + \frac{C_2^{}}{N_{\rm c}^{}} \,\,, \hspace{2em}
a_1^{\rm LR,RL} \,\,=\,\, C_1^{\rm LR,RL} + \frac{C_2^{\rm LR,RL}}{N_{\rm c}^{}} \,\,.
\end{eqnarray}
The presence of the second term in the \,$\bar B\to\eta_c^{}\bar K$\, amplitude offers
the possibility of $\sin\bigl(2\beta^{\rm eff}\bigr)$ in this decay mode being different from that in
\,$\bar B\to J/\psi\bar K$.\,  Defining
\begin{eqnarray}
r_\kappa^{}(\mu) \,\,=\,\,
\frac{\bigl(a_1^{\rm LR}(\mu)-a_1^{\rm RL}(\mu)\bigr)m_{\eta_c^{}}^2}
     {a_1^{}(\mu)\,m_c^{}(\mu)\,\bigl(m_b^{}(\mu)-m_s^{}(\mu)\bigr)} \,\,,
\end{eqnarray}
we then obtain to first order in $\kappa$
\begin{eqnarray}  \label{beta}
\beta^{\rm eff}_{\eta_c^{}K} \,\,=\,\, \beta^{\rm eff}_{\psi K}
+ {\rm arg}\bigl(1+r_\kappa^{}\bigr)
\,\,\simeq\,\,  \beta^{\rm eff}_{\psi K} + {\rm Im}\,r_\kappa^{} \,\,.
\end{eqnarray}
Taking \,$\mu=m_b^{}=4.2{\rm\,GeV}$\, and \,$N_{\rm c}^{}=3$,\, we find \,$a_1^{}(\mu)=0.076$\,
and  \,$r_\kappa^{}(\mu)\simeq 20\bigl(\kappa_{cb}^{\rm R}-\kappa_{cs}^{\rm R*}\bigr)$.\,
Since the experimental numbers in Eq.~(\ref{sin2b}) imply
\begin{eqnarray} \label{betad}
\beta_{\eta_c^{}K}^{\rm eff} \,\,=\,\, 0.60 \pm 0.23 \,\,, \hspace{5ex}
\beta_{\psi K}^{\rm eff} \,\,=\,\, 0.358 \pm 0.017 \,\,,
\end{eqnarray}
in view of Eq.~(\ref{beta}) we can then impose \,$-0.005\le{\rm Im}\,r_\kappa^{}(\mu)\le 0.4$,\,
which leads to the bound
\,$-2.5\times10^{-4}\le{\rm Im}\bigl(\kappa_{cb}^{\rm R}+\kappa_{cs}^{\rm R}\bigr) \le 0.02$.\,

It is well known, however, that this naive factorization procedure fails to reproduce the experimental
branching ratios, which can be better fit with  \,$N_{\rm c}\simeq 2$\,~\cite{nceff}.
Using this value we obtain instead
\,$r_\kappa^{}(\mu)\simeq8\bigl(\kappa_{cb}^{\rm R}-\kappa_{cs}^{\rm R*}\bigr)$.\,
This would increase the upper bound for  \,${\rm Im}\bigl(\kappa_{cb}^{\rm R}+\kappa_{cs}^{\rm R}\bigr)$\,
above by about a factor of two, within the intrinsic uncertainty of our calculation,
\begin{eqnarray} \label{im(cb+cs)}
-5\times10^{-4} \,\,\le\,\, {\rm Im}\bigl(\kappa_{cb}^{\rm R}+\kappa_{cs}^{\rm R}\bigr) \,\,\le\,\, 0.04 \,\,.
\end{eqnarray}

\section{Constraints from dipole penguin operators}

We turn next to constraints from the magnetic-penguin transitions  \,$d\to d'\gamma$\,  and
\,$d\to d'g$.\,  The specific processes we discuss are \,$b\to s\gamma$,\, \,$d\to s\gamma$,
the $CP$-violation parameters $\epsilon$ and $\epsilon'$ in kaon mixing and decay, and
hyperon $CP$ violation.

\subsection{\boldmath$b\to s\gamma$}

Including the SM contribution, the effective Hamiltonian for \,$b\to s\gamma$\, is
\begin{eqnarray}
{\cal H}_{b\to s\gamma}^{} \,\,=\,\,
\frac{-e G_{\rm F}^{}}{4\sqrt2\,\pi^2}\sum_{q=u,c,t} \bar s\sigma^{\mu\nu}
\bigl(F_{\rm L}^q P_{\rm L}^{}+F_{\rm R}^q P_{\rm R}^{}\bigr)b\,F_{\mu\nu}^{} \,\,,
\end{eqnarray}
where to ${\cal O}(\kappa)$
\begin{eqnarray}
\begin{array}{c}   \displaystyle
F_{\rm L}^q \,\,=\,\, V_{qs}^*V_{qb}^{} \Bigl[
\bigl(1+\kappa_{qs}^{\rm L*}+\kappa_{qb}^{\rm L}\bigr)\, m_s^{}\, F_0^{\rm SM}\big(x_q^{}\bigr) +
\kappa_{qs}^{\rm R*}\, m_q^{}\, F_0^{}\big(x_q^{}\bigr) \Bigr] \,\,,
\vspace{2ex} \\   \displaystyle
F_{\rm R}^q \,\,=\,\, V_{qs}^*V_{qb}^{} \Bigl[
\bigl(1+\kappa_{qs}^{\rm L*}+\kappa_{qb}^{\rm L}\bigr)\, m_b^{}\, F_0^{\rm SM}\big(x_q^{}\bigr) +
\kappa_{qb}^{\rm R}\, m_q^{}\, F_0^{}\big(x_q^{}\bigr) \Bigr] \,\,,
\end{array}
\end{eqnarray}
following from Eqs.~(\ref{Hdd'f}) and~(\ref{FLR}).
The corresponding expressions for \,$b\to s g$\, are similar in form and follow from
Eqs.~(\ref{Hdd'g}) and~(\ref{GLR}).

The experimental data on \,$b\to s\gamma$\, have been found to impose very strong constraints on
$\kappa_{tb,ts}^{}$, limiting them to below the few-percent level~\cite{Fujikawa:1993zu}.
Since  \,$V_{cs}^*V_{cb}^{}\simeq-V_{ts}^*V_{tb}^{}$\,  and  \,$m_t^{}\gg m_c^{}$,\, the preceding
equations indicate that, if all the $\kappa$'s were comparable in size, the top contributions
would be larger than the charm ones by almost two orders of magnitude.
All this means that \,$b\to s\gamma$\,  offers relatively weak constraints on $\kappa_{cb,cs}^{}$,
with upper bounds at the level of~${\cal O}(1)$.

\subsection{\boldmath$s\to d\gamma$}

In an analogous manner, the anomalous couplings contribute to the short-distance transition
\,$s\to d\gamma$,\, but in this case the charm contribution is expected to be more important than
the top one.  At lower energies, $C_\gamma$ and~$C_g$ mix because of QCD corrections.
At  $\,\mu=1\,\rm GeV\,$  we have~\cite{Cho:1993zb}
\begin{eqnarray}
C_\gamma^{}(\mu) \,\,=\,\, \bar\eta^8\, C_\gamma^{}(m_W^{})
\,+\, \mbox{$\frac{8}{3}$}\bigl(\bar\eta^7-\bar\eta^8\bigr)\, C_g^{}(m_W^{}) \,\,,
\end{eqnarray}
where
\begin{eqnarray} \label{heta}
\bar\eta \,\,=\,\,
\Biggl(\frac{\alpha_{\rm s}^{}(m_W^{})}{\alpha_{\rm s}^{}(m_b^{})}\Biggr)^{\!\!2/23}
\Biggl(\frac{\alpha_{\rm s}^{}(m_b^{})}{\alpha_{\rm s}^{}(m_c^{})}\Biggr)^{\!\!2/25}
\Biggl(\frac{\alpha_{\rm s}^{}(m_c^{})}{\alpha_{\rm s}^{}(\mu)}\Biggr)^{\!\!2/27} \,\,.
\end{eqnarray}
Numerically, keeping only the charm contributions yields
\begin{eqnarray}
C_{\gamma}^\pm(\mu) \,\,=\,\,
(-38+0.023\,i)\bigl(\kappa_{cs}^{\rm R}\pm\kappa_{cd}^{\rm R*}\bigr)\times10^{-7}\, {\rm GeV}^{-1} \,\,.
\end{eqnarray}

Hyperon and kaon radiative-weak decays provide the relevant constraints, the former being somewhat
stronger and yielding~\cite{He:1999ik}
\begin{eqnarray}
\frac{\bigl|C_\gamma^+(\mu)\bigr|}{8\pi^2\,G_{\rm F}^{}} \,\,\le\,\, 12{\rm\,MeV} \,\,.
\end{eqnarray}
This translates into
\begin{eqnarray} \label{kcdcs}
\bigl|\kappa_{cd}^{\rm R*}+\kappa_{cs}^{\rm R}\bigr| \,\,\le\,\, 3 \,\,,
\end{eqnarray}
which is a very weak bound compared to Eqs.~(\ref{recd}) and~(\ref{recs}).

\subsection{\boldmath$\epsilon$ and $\epsilon'$}

The gluonic dipole operators contribute to the $CP$-violation parameters $\epsilon$ and
$\epsilon'$ in kaon mixing and decay, respectively.
Since $Q^+_g$ is parity conserving, it contributes to $\epsilon$ via long-distance
effects~\cite{Donoghue:1985ae,He:1999bv}.
Being parity violating, $Q^-_g$ contributes to $\epsilon'$.
One finds~\cite{He:1999bv,Buras:1999da,Tandean:2003fr}
\begin{eqnarray} \label{e}
(\epsilon)_\kappa^{} \,\,=\,\, -2.3\times10^5\,{\rm GeV}\, B_\epsilon^{}\, {\rm Im}\,C_g^+(\mu) \,\,,
\hspace{5ex}
\biggl(\frac{\epsilon'}{\epsilon}\biggr)_{\!\!\kappa}  \,\,=\,\,
4.4\times10^5\,{\rm GeV}\, B_{\epsilon'}^{}\,\, {\rm Im}\, C_g^-(\mu) \,\,,
\end{eqnarray}
where the contributions of the anomalous charm couplings to $C_g^\pm$ are
\begin{eqnarray} \label{cg1}
C_g^\pm(\mu) \,\,=\,\, \bar\eta^7\, C_g^\pm(m_W^{}) \,\,=\,\,
(-21+0.013\,i)\bigl(\kappa_{cs}^{\rm R}\pm\kappa_{cd}^{\rm R*}\bigr)\times10^{-7}\, {\rm GeV}^{-1}
\end{eqnarray}
for $\,\mu=1\,\rm GeV\,$  and the hadronic uncertainties are represented by
\begin{eqnarray} \label{bebep}
0.2 \,\,\le\,\, |B_\epsilon^{}| \,\,\le\,\, 1 \,\,, \hspace{7ex}
0.5 \,\,\le\,\, |B_{\epsilon'}^{}| \,\,\le\,\, 2 \,\,.
\end{eqnarray}
The experimental data are \,$|\epsilon|=(2.229\pm0.012)\times10^{-3}$\, and
\,${\rm Re}(\epsilon'/\epsilon)=(1.65\pm0.26)\times10^{-3}$\,~\cite{pdg}.
The SM predicts  \,$|\epsilon|_{\rm SM}=\bigl(2.06^{+0.47}_{-0.53}\bigr)\times10^{-3}$\,~\cite{ckmfit},
but for $\epsilon'$ the SM calculation still involves a large uncertainty~\cite{Eeg:2008kf}.
Consequently, we require that
\begin{eqnarray}   \label{eep}
|\epsilon|_\kappa^{} \,\,<\,\, 0.7\times10^{-3} \,\,, \hspace{7ex}
\biggl(\frac{\epsilon'}{\epsilon}\biggr)_{\!\!\kappa}  \,\,<\,\, 1.7\times10^{-3} \,\,.
\end{eqnarray}
The resulting constraints on $\kappa_{cd,cs}^{\rm R}$ are complicated and will be presented in
Fig.~\ref{kplots} in Sec.~\ref{final}.

There are other loop-generated operators contributing to~$\epsilon$, and hence they provide more
constraints on the anomalous charm couplings.
These operators will be discussed in Sec.~\ref{kkmix}

\subsection{Hyperon nonleptonic decays}

Hyperon decays provide an additional environment to study $CP$-violating \,$|\Delta S|=1$\,
interactions.
The main observable of interest in this case is the $CP$-violating asymmetry
\,$A=\bigl(\alpha+\bar\alpha\bigr)/\bigl(\alpha-\bar\alpha\bigr)$,\, where
$\alpha$ is a decay parameter in the decay of a hyperon into another baryon and a spinless meson
and $\bar\alpha$ is the corresponding parameter in the antiparticle process~\cite{Donoghue:1986hh}.

Experimentally, a preliminary value \,$A_{\Lambda\Xi}^{\rm exp}=(-6\pm3)\times10^{-4}$\, for $A$
measured in the decay chain  \,$\Xi\to\Lambda\pi\to p\pi\pi$\, has recently been reported~\cite{hypercp}.
The SM prediction is
\,$\bigl|A_{\Xi\Lambda}^{\rm SM}\bigr|\lesssim 5\times 10^{-5}$\,~\cite{Donoghue:1986hh},
which is an order of magnitude smaller than the central value of the measurement.
Since this is only a 2-sigma disagreement, it is premature to attribute it to new physics.
However, this difference can also be used to constrain the anomalous charm couplings.

The contribution of the gluonic dipole operators to the asymmetry $A_{\Lambda\Xi}^{}$  has been
estimated in Ref.~\cite{Tandean:2003fr}.
The result can be written as
\begin{eqnarray}
\bigl(A_{\Xi\Lambda}^{}\bigr)_\kappa \,\,=\,\,
10^5\, B_+^{}\,{\rm Im}\,C_g^+(\mu) \,+\, 10^5\, B_-^{}\,{\rm Im}\,C_g^-(\mu)  \,\,,
\end{eqnarray}
where  $C_g^\pm(\mu)$ due to the anomalous charm couplings are given in Eq.~(\ref{cg1})  and
\begin{eqnarray} \label{b+b-}
-1.4 \,\,\le\,\, B_+^{} \,\,\le\,\, 0.5 \,\,, \hspace{7ex} -0.9\,\,\le\,\, B_-^{} \,\,\le\,\, 1.3
\end{eqnarray}
reflect the hadronic uncertainties.
The preliminary HyperCP result above suggests that
\begin{eqnarray} \label{ax}
-9\times10^{-4} \,\,<\,\, \bigl(A_{\Xi\Lambda}^{}\bigr)_\kappa \,\,<\,\, -3\times10^{-4} \,\,.
\end{eqnarray}
The resulting constraints are shown in Fig.~\ref{kplots} in Sec.~\ref{final}.

\section{Electric dipole moment of neutron}

The flavor-conserving counterparts of the magnetic-dipole operators discussed above contribute
to the neutron EDM.
The latter is described by the effective Lagrangian
\begin{eqnarray}
{\cal L}_{n\rm edm}^{}  \,\,=\,\,
-\mbox{$\frac{i}{2}$}\, d_n^{}\, \bar n\,\sigma^{\mu\nu}\gamma_5^{}\,n\, F_{\mu\nu}^{} \,\,.
\end{eqnarray}
The dipole moments $d_d^{\rm edm}$ and $d_d^{\rm cdm}$ of the $d$ quark in Eq.~(\ref{dedm})
contribute to $d_n^{}$.
Using the valence quark model, we have~\cite{He:1989xj}
\begin{eqnarray} \label{dnd}
d_n^{(d)} \,\,=\,\, \mbox{$\frac{4}{3}$}\,d_d^{\rm edm}(\mu) +
\mbox{$\frac{4}{9}$} \,e\,d_d^{\rm cdm}(\mu) \,\,,
\end{eqnarray}
where at \,$\mu=1$\,GeV\,
\begin{eqnarray} \label{dedm'}
\begin{array}{c}   \displaystyle
d_d^{\rm edm}(\mu) \,\,=\,\, \bar\eta^8\, d_d^{\rm edm}(m_W^{})
\,+\, \mbox{$\frac{8}{3}$}\bigl(\bar\eta^7-\bar\eta^8\bigr)\, e\, d_d^{\rm cdm}(m_W^{}) \,\,,
\vspace{2ex} \\   \displaystyle
d_d^{\rm cdm}(\mu) \,\,=\,\, \bar\eta^7\, d_d^{\rm cdm}(m_W^{}) \,\,,
\end{array}
\end{eqnarray}
with  $\bar\eta$ being given in Eq.~(\ref{heta}).
The anomalous charm contribution is then
\begin{eqnarray}
d_n^{(d)} \,\,=\,\, 6.9\,{\rm Im}\,\kappa_{cd}^{\rm R}\, \times 10^{-22}\, e{\rm\,cm}  \,\,.
\end{eqnarray}
Similarly, the electric and color dipole-moments of the $s$ quark produced by the anomalous
charm couplings are
\begin{eqnarray}
\begin{array}{c}   \displaystyle
d_s^{\rm edm}(\mu) \,\,=\,\,
\frac{-e G_{\rm F}^{}}{2\sqrt2\,\pi^2} \bigl|V_{cs}^{}\bigr|^2\,
{\rm Im}\,\kappa_{cs}^{\rm R}\, m_c^{}\, \Bigl[ \bar\eta^8\, F_0^{}(x_c^{})
\,+\, \mbox{$\frac{8}{3}$}\bigl(\bar\eta^7-\bar\eta^8\bigr)\, G_0^{}(x_c^{}) \Bigr] \,\,,
\vspace{2ex} \\   \displaystyle
d_s^{\rm cdm}(\mu) \,\,=\,\,
-\bar\eta^7\, \frac{g_{\rm s}^{} G_{\rm F}^{}}{2\sqrt2\,\pi^2} \bigl|V_{cs}^{}\bigr|^2\,
{\rm Im}\,\kappa_{cs}^{\rm R}\, m_c^{}\, G_0^{}(x_c^{}) \,\,,
\end{array}
\end{eqnarray}
and so their contribution to $d_n^{}$ is given by
\begin{eqnarray} \label{dns}
d_n^{(s)} \,\,=\,\, B_{\rm e}^{}\, d_s^{\rm edm}(\mu) + B_{\rm c}^{}\,e\,d_s^{\rm cdm}(\mu)
\,\,=\,\,
\bigl(82\,B_{\rm e}^{}+46\,B_{\rm c}^{}\bigr)\,{\rm Im}\,\kappa_{cs}^{\rm R}\,
\times 10^{-22}\,e{\rm\,cm} \,\,,
\end{eqnarray}
where  \,$-0.35\le B_{\rm e}^{}\le-0.01$\,  and  \,$0.01\le B_{\rm c}^{}\le 0.26$\,
reflect the wide range of estimates for $d_n^{(s)}$ in the literature~\cite{Dib:2006hk},
in contrast to those for $d_n^{(d)}$.
The combined contribution of $d_n^{(d,s)}$ is then
\begin{eqnarray} \label{dn}
(d_n^{})_\kappa^{} \,\,=\,\, d_n^{(d)} + d_n^{(s)} \,\,=\,\,
\Bigl(0.69\,{\rm Im}\,\kappa_{cd}^{\rm R} \,+\, B_n^{}\,{\rm Im}\,\kappa_{cs}^{\rm R} \Bigr)\times
10^{-21}\, e{\rm\,cm}  \,\,,
\end{eqnarray}
where
\begin{eqnarray} \label{bn}
-2.8 \,\,\le\,\, B_n^{} \,\,\le\,\, +1.1 \,\,.
\end{eqnarray}
From the experimental bound  \,$|d_n^{}|_{\rm exp}^{}<2.9\times10^{-26}\,e$\,cm\, at
90\%\,C.L.~\cite{pdg}, we will impose the bound
\begin{eqnarray} \label{dn_bound}
|d_n^{}|_\kappa^{} \,\,<\,\, 2.9\times10^{-26}\,e{\rm\,cm}
\end{eqnarray}
in Sec.~\ref{final} to restrict the anomalous couplings further.

The $s$-quark dipole moments $d_s^{\rm edm}$ and $d_s^{\rm edm}$ above also contribute to
the EDM of the $\Lambda$ hyperon and therefore may be constrained directly by experiment.
However, the experimental limit,
\,$d_\Lambda^{}=(-3.0\pm 7.4)\times10^{-17}\,e$\,cm\,~\cite{Pondrom:1981gu} or
\,$d_\Lambda^{}<1.5\times10^{-16}\,e$\,cm\, at 90\% C.L.~\cite{pdg},
is very weak compared to Eq.~(\ref{dn_bound}) and hence will not be used for constraining
the couplings.

\section{Other loop constraints}

In this section we explore several other processes where the anomalous charm couplings can
contribute via penguin and box diagrams.

\subsection{\boldmath$K^+\to\pi^+\nu\bar\nu$}

To quantify the contribution of the anomalous charm couplings to this mode, it is convenient to
compare it with the dominant contribution in the SM.
The latter comes from the top loop and is given by~\cite{Buchalla:1995vs,Inami:1980fz}
\begin{eqnarray}
{\cal M}_{\rm SM}^{}\bigl(K^+\to\pi^+\nu\bar\nu\bigr) \,\,=\,\,
\frac{G_{\rm F}^{}}{\sqrt2}\, \frac{\alpha}{2\pi\,\sin^2\theta_{\rm W}^{}}\,\,
V_{td}^{}V_{ts}^*\, X_0^{}\bigl(x_t^{}\bigr)\, \langle\pi^+|\bar s\gamma_\mu^{}d|K^+\rangle\,
\bar\nu\gamma^\mu(1-\gamma_5^{})\nu \,\,,
\end{eqnarray}
following from Eq.~(\ref{dd2nn_sm}), without QCD corrections.
It is also convenient to neglect the masses of the leptons associated with the neutrinos
in the new contribution, as in Eq.~(\ref{hnunu}), so that we can work with just one of them.
The total amplitude can thus be written in terms of the SM amplitude above as
\begin{eqnarray}
{\cal M}(K^+\to\pi^+\nu\bar\nu) \,\,=\,\,
(1+\delta)\,{\cal M}_{\rm SM}^{}\bigl(K^+\to\pi^+\nu\bar\nu\bigr) \,\,,
\end{eqnarray}
where to linear order in $\kappa$
\begin{eqnarray}
\delta \,\,=\,\, \frac{V_{cd}^{}V_{cs}^*}{V_{td}^{}V_{ts}^*}\,
\frac{\bigl(\kappa_{cd}^{\rm L}+\kappa_{cs}^{\rm L*}\bigr)
\bigl[-3\,\ln\bigl(\Lambda/m_W^{}\bigr)+4X_0^{}\bigl(x_c^{}\bigr)\bigr]}{4 X\bigl(x_t^{}\bigr)} \,\,.
\end{eqnarray}
In the above expression we have used \,$X\bigl(x_t^{}\bigr)\simeq1.4$\,  instead of
$X_0^{}\bigl(x_t^{}\bigr)$ in the denominator to incorporate the  QCD corrections in
the SM~\cite{Buras:2006gb}.  The SM prediction for the branching ratio is
\,${\cal B}_{\rm SM}^{}(K^+\to\pi^+\nu\bar\nu)=(8.5\pm0.7)\times10^{-11}$\,~\cite{Buras:2006gb},
to be compared with its experimental value
\,${\cal B}_{\rm exp}^{}=\bigl(1.73_{-1.05}^{+1.15}\bigr)\times10^{-10}$\,~\cite{Artamonov:2008qb}.
Accordingly, we require  \,$-0.2\le{\rm Re}\,\delta\le 1$,\,  which translates into
\begin{eqnarray} \label{k2pnn}
-2.5\times10^{-4} \,\,\le\,\, -{\rm Re}\bigl(\kappa^{\rm L}_{cd}+\kappa^{\rm L}_{cs}\bigr) \,+\,
0.42\,{\rm Im}\bigl(\kappa^{\rm L}_{cd}-\kappa^{\rm L}_{cs}\bigr) \,\,\le\,\, 1.3\times10^{-3} \,\,.
\end{eqnarray}

It is interesting to compare the anomalous charm contribution to the SM charm contribution.
Their ratio is
\begin{eqnarray}
\frac{{\cal M}_\kappa^{}\bigl(K^+\to\pi^+\nu\bar\nu\bigr)}
     {{\cal M}_{\rm SM}^{(c)}\bigl(K^+\to\pi^+\nu\bar\nu\bigr)} \,\,=\,\,
\frac{\bigl(\kappa_{cd}^{\rm L}+\kappa_{cs}^{\rm L*}\bigr)
\bigl[-3\,\ln\bigl(\Lambda/m_W^{}\bigr)+4X_0^{}\bigl(x_c^{}\bigr)\bigr]}{4X_{\rm NL}^{}} \,\,,
\end{eqnarray}
where  \,$6\times10^{-4}\lesssim X_{\rm NL}^{}\lesssim1\times10^{-3}$\, incorporates QCD corrections
and lepton-mass dependence~\cite{Buchalla:1995vs}.
With the \,$\kappa^{\rm L}_{cd}+\kappa^{\rm L}_{cs}$\, at the upper end of the range above,
the two contributions are similar in size.
This implies that the current experimental situation admits a~$100\%$  uncertainty in the charm
contribution to the branching ratio, much larger than the theoretical uncertainty within the SM.

Constraints on the anomalous couplings can also be extracted from the related $B$-meson modes,
\,$B\to X\nu\bar\nu$,\, but the resulting bounds are about three orders of magnitude weaker
due to unfavorable CKM angles.  Experimentally, only upper limits for their decay rates are
currently available~\cite{hfag}. For these reasons, we do not discuss them further.

\subsection{\boldmath$K_L\to\mu^+\mu^-$}

The dominant part of the short-distance contribution to the SM amplitude for \,$K^0\to\mu^+\mu^-$\,
is again induced by the top loop and can be expressed as~\cite{Buchalla:1995vs,Inami:1980fz}
\begin{eqnarray}
{\cal M}_{\rm SM}^{\rm SD}\bigl(K^0\to\mu^+\mu^-\bigr) \,\,=\,\,
-\frac{G_{\rm F}^{}}{\sqrt2}\, \frac{\alpha}{2\pi\,\sin^2\theta_{\rm W}^{}}\, V_{td}^{}V_{ts}^*\,
Y_0^{}\bigl(x_t^{}\bigr)\, \langle0|\bar s\gamma^\sigma\gamma_5^{}d|K^0\rangle\,
\bar \mu\gamma_\sigma^{}\gamma_5^{}\mu \,\,,
\end{eqnarray}
from Eq.~(\ref{hll}).
Combining this with the anomalous charm contribution in Eq.~(\ref{hll}), we arrive at
the total short-distance amplitude
\begin{eqnarray}
{\cal M}_{\rm SD}^{}\bigl(K_L^{}\to\mu^+\mu^-\bigr) \,\,=\,\,
(1+\delta')\, {\cal M}_{\rm SM}^{\rm SD}\bigl(K_L^{}\to\mu^+\mu^-\bigr)  \,\,,
\end{eqnarray}
where to linear order in $\kappa$
\begin{eqnarray}
\delta' \,\,=\,\,
\frac{{\rm Re}\bigl[V_{cd}^*V_{cs}^{}\,\bigl(\kappa_{cs}^{\rm L}+\kappa_{cd}^{\rm L*}\bigr)\bigr]
\bigl[-3\,\ln\bigl(\Lambda/m_W^{}\bigr)+4 Y_0^{}\bigl(x_c^{}\bigr)\bigr]}
{4\,{\rm Re}\bigl(V_{td}^*V_{ts}^{}\bigr)\, Y\bigl(x_t^{}\bigr)} \,\,,
\end{eqnarray}
with  \,$Y\bigl(x_t^{}\bigr)\simeq0.95$\, being the QCD-corrected value of
$Y_0^{}\bigl(x_t^{}\bigr)$~\cite{Gorbahn:2006bm}.
Since the measured branching ratio,
\,${\cal B}\bigl(K_L^{}\to\mu^+\mu^-\bigr)=(6.84\pm0.11)\times10^{-9}$\,~\cite{pdg},  is almost
saturated by the absorptive part of the long-distance contribution,
\,${\cal B}_{\rm abs}^{}=(6.64\pm0.07)\times10^{-9}$\,~\cite{Littenberg:2008zz}, the difference
between them suggests the allowed room for new physics,  \,${\cal B}_{\rm NP}^{}\lesssim3.8\times10^{-10}$,\,
the upper bound being about one half of the SM short-distance contribution,
\,${\cal B}_{\rm SM}^{\rm SD}=(7.9\pm1.2)\times 10^{-10}$\,~\cite{Gorbahn:2006bm}.
Consequently, we demand \,$|\delta'|\le 0.2$,\, which implies
\begin{eqnarray} \label{k2ll}
\Bigl|{\rm Re}\bigl(\kappa_{cs}^{\rm L}+\kappa_{cd}^{\rm L}\bigr) \,+\,
6\times10^{-4}\,{\rm Im}\bigl(\kappa_{cs}^{\rm L}-\kappa_{cd}^{\rm L}\bigr)\Bigr|
\,\,\le\,\, 1.5\times 10^{-4} \,\,.
\end{eqnarray}

One could also carry out a similar analysis as above for  \,$B\to\ell^+\ell^-$,\,
but the CKM angles in that case are such that the constraints would be much weaker.
In addition, only experimental bounds on the rates are currently available~\cite{hfag}.

\subsection{\boldmath$K$-$\bar K$ mixing\label{kkmix}}

The matrix element $M_{12}^{}$ for $K^0$-$\bar K^0$ mixing is defined by~\cite{Buchalla:1995vs}
\begin{eqnarray}
2m_K^{}\, M_{12}^K \,\,=\,\,
\bigl\langle K^0\bigr| {\cal H}_{d\bar s\to\bar d s}^{} \bigl|\bar K^0\bigr\rangle
\end{eqnarray}
where the effective Hamiltonian ${\cal H}_{d\bar s\to\bar d s}$ consists of SM and
new-physics terms.
For the latter, the contribution of the anomalous charm couplings can be derived from
Eq.~(\ref{hbox}), where the last line is negligible compared to the second because of
the smallness of $x_c^{}$.  Thus
\begin{eqnarray} \label{mk}
M_{12}^{K,\kappa} &=& \frac{G_{\rm F}^2\,m_W^2}{24\pi^2}\, f_K^2 m_K^{}\, \lambda_c^{ds} \left[
\bar\eta^3 B_K^{}\, \bigl(\kappa_{cd}^{\rm L*}+\kappa_{cs}^{\rm L}\bigr)
\left( -\lambda_t^{ds}\, x_t^{}\,\ln\frac{\Lambda^2}{m_W^2}
- \sum_q\lambda_q^{ds}\, {\cal B}_1^{}\bigl(x_q^{},x_c^{}\bigr) \right) \right.
\nonumber \\ && \hspace*{20ex} +\, \left.
\frac{\bar\eta^{3/2} B_K^{}\, m_K^2}{\bigl(m_d^{}+m_s^{}\bigr)^2}\,
\kappa_{cd}^{\rm R*}\kappa_{cs}^{\rm R} \left( \lambda_t^{ds}\,x_t^{}\,\ln\frac{\Lambda^2}{m_W^2}
+ \sum_q\lambda_q^{ds}\, {\cal B}_2^{}\bigl(x_q^{},x_c^{}\bigr) \right) \right]
\nonumber \\ &=&
-(0.090+0.031\,i)\,{\rm ps}^{-1}\, \bigl(\kappa_{cd}^{\rm L*}+\kappa_{cs}^{\rm L}\bigr) \,+\,
(2.1+0.58\,i)\,{\rm ps}^{-1}\, \kappa_{cd}^{\rm R*}\kappa_{cs}^{\rm R} \,\,,
\end{eqnarray}
where \,$\lambda_q^{ds}=V_{qd}^*V_{qs}^{}$,\,  we have included QCD-correction factors at
leading order with $\bar\eta$ given in Eq.~(\ref{heta}),  \,$m_d^{}(\mu)+m_s^{}(\mu)=142$\,MeV\,
at \,$\mu=1$\,GeV,\, and the other parameters can be found in Appendix~\ref{me}.
Evidently, the inclusion of the second term in $M_{12}^{K,\kappa}$, albeit quadratic in
$\kappa^{\rm R}$, is important as it receives large chiral and QCD enhancement with respect to
the first term.

Now, the difference  $\Delta M_K^{}$  between the $K_L^{}$ and $K_S^{}$ masses is related to
\,$M_{12}^K=M_{12}^{K,\rm SM}+M_{12}^{K,\kappa}$\,  by
\,$\Delta M_K^{}=2\,{\rm Re}\,M_{12}^K+\Delta M_K^{\rm LD}$,\,  the long-distance term
$\Delta M_K^{\rm LD}$ being sizable~\cite{Buchalla:1995vs}.
Since the LD part suffers from significant uncertainties, we constrain the anomalous couplings
by requiring that their contribution to $\Delta M_K^{}$ be less than the largest SM contribution,
which comes from the charm loop and is given by
\begin{eqnarray}
M_{12}^{K,\rm SM} \,\,\simeq\,\, \frac{G_{\rm F}^2\,m_W^2}{12\pi^2}\, f_K^2 m_K^{} B_K^{}\,
\eta_{cc}^{}\bigl(\lambda_c^{ds}\bigr)^2\,S_0^{}\bigl(x_c^{}\bigr) \,\,,
\end{eqnarray}
with the parameter values in Appendix~\ref{me}.
The result is
\begin{eqnarray} \label{kk1}
\bigl|0.043\, {\rm Re}\bigl(\kappa_{cd}^{\rm L}+\kappa_{cs}^{\rm L}\bigr)
+ 0.015\,{\rm Im}\bigl(\kappa_{cd}^{\rm L}-\kappa_{cs}^{\rm L}\bigr)
\,-\, {\rm Re}\bigl(\kappa_{cd}^{\rm R*}\kappa_{cs}^{\rm R}\bigr)
+ 0.28\,{\rm Im}\bigl(\kappa_{cd}^{\rm R*}\kappa_{cs}^{\rm R}\bigr) \bigr|
\,\,\le\,\, 8.5\times10^{-4} \,\,. \;\;\;
\end{eqnarray}

A complementary constraint on the couplings can be obtained from the $CP$-violation
parameter~$\epsilon$.  Its magnitude is related to $M_{12}^K$ by~\cite{Buchalla:1995vs}
\begin{eqnarray}
|\epsilon| \,\,\simeq\,\, \frac{\bigl|{\rm Im}\,M_{12}^K\bigr|}{\sqrt2\,\Delta M_K^{\rm exp}} \,\,,
\end{eqnarray}
where  \,$\Delta M_K^{\rm exp}=(3.483\pm0.006)\times10^{-15}$\,GeV\,~\cite{pdg} and the small
term containing the $CP$-violating phase in the \,$K\to\pi\pi$\, amplitude has been dropped.
Since  \,$|\epsilon|_{\rm exp}^{}=(2.229\pm0.012)\times10^{-3}$\,~\cite{pdg} and
\,$|\epsilon|_{\rm SM}=\bigl(2.06^{+0.47}_{-0.53}\bigr)\times10^{-3}$\,~\cite{ckmfit}, we again
demand  \,$|\epsilon|_\kappa^{}<0.7\times10^{-3}$\,  for the contribution in Eq.~(\ref{mk}).
This translates into
\begin{eqnarray} \label{kk2}
\bigl|0.015\, {\rm Re}\bigl(\kappa_{cs}^{\rm L}+\kappa_{cd}^{\rm L}\bigr)
+ 0.043\,{\rm Im}\bigl(\kappa_{cs}^{\rm L}-\kappa_{cd}^{\rm L}\bigr)
\,-\, 0.28\,{\rm Re}\bigl(\kappa_{cd}^{\rm R*}\kappa_{cs}^{\rm R}\bigr)
- {\rm Im}\bigl(\kappa_{cd}^{\rm R*}\kappa_{cs}^{\rm R}\bigr) \bigr|
\,\le\, 2.5\times10^{-6} \,\,. \;\;\;
\end{eqnarray}

\subsection{\boldmath$B_d$-$\bar B_d$ mixing}

In the SM the matrix element $M_{12}^d$ for $B_d^0$-$\bar B_d^0$ mixing is dominated by the top-loop
contribution and given by~\cite{Buchalla:1995vs}
\begin{eqnarray}
M_{12}^{d,\rm SM} \,\,=\,\,
\frac{\bigl\langle B_d^0\bigr| {\cal H}_{d\bar b\to\bar d b}^{\rm SM} \bigl|\bar B_d^0\bigr\rangle}
     {2 m_{B_d}^{}}
\,\,\simeq\,\,
\frac{G_{\rm F}^2\,m_W^2}{12\pi^2}\, f_{B_d}^2 m_{B_d}^{}\, \eta_B^{}B_{B_d}^{}\,
\bigl(V_{tb}^{}V_{td}^*\bigr)^2\,S_0^{}\bigl(x_t^{}\bigr)
\end{eqnarray}
following from Eq.~(\ref{dd2dd_sm}), where a QCD-correction factor $\eta_B^{}$ has been
included and the parameter values can be found in Appendix~\ref{me}.
In contrast, since $V_{cd}^*V_{cb}^{}$ and $V_{td}^*V_{tb}^{}$ are comparable in size,
the anomalous couplings of charm and top may produce similar effects on $M_{12}^d$, as can be
inferred from Eq.~(\ref{hbox}).
The anomalous top contributions to $B$ mixing having been studied before~\cite{AbdElHady:1997eu},
we switch them off and get, to linear order in $\kappa$,
\begin{eqnarray}
M_{12}^{d,\kappa} &=&
\frac{G_{\rm F}^2 m_W^2}{24\pi^2}\, f_{B_d}^2 m_{B_d}^{}\, \eta_B^{}B_{B_d}^{}\,
\lambda_c^{db}\bigl(\kappa_{cb}^{\rm L}+\kappa_{cd}^{\rm L*}\bigr)
\Biggl( -\lambda_t^{db}\, x_t^{}\,\ln\frac{\Lambda^2}{m_W^2} -
\sum_q\lambda_q^{db}\, {\cal B}_1^{}\bigl(x_q^{},x_c^{}\bigr) \Biggr)
\nonumber \\ &=&
(1.6+0.63\,i)\,{\rm ps}^{-1}\, \bigl(\kappa_{cb}^{\rm L}+\kappa_{cd}^{\rm L*}\bigr) \,\,,
\end{eqnarray}
where \,$\lambda_q^{db}=V_{qd}^*V_{qb}^{}$\,  and we have neglected the terms quadratic in
$\kappa^{\rm R}$ because their quark operators do not get as much chiral and QCD enhancement
as those in the kaon-mixing case.

The difference  $\Delta M_d^{}$  between the masses of the heavy and light mass-eigenstates
is related to  \,$M_{12}^d=M_{12}^{d,\rm SM}+M_{12}^{d,\kappa}$\,  by
\,$\Delta M_d^{}=2\bigl|M_{12}^d\bigr|$\,~\cite{Buchalla:1995vs}.
The measured value  \,$\Delta M_d^{\rm exp}=(0.507\pm0.005)\,{\rm ps}^{-1}$\,~\cite{pdg}
agrees with the SM prediction,
\,$\Delta M_d^{\rm SM}=\bigl(0.563^{+0.068}_{-0.076}\bigr)\,{\rm ps}^{-1}$\,~\cite{ckmfit}.
In the presence of the anomalous couplings, these numbers are related by
\begin{eqnarray} \label{dd}
\Delta M_d^{\rm exp}\,\,=\,\, \Delta M_d^{\rm SM}\,\bigl|1+\delta_d^{}\bigr| \,\,, \hspace{5ex}
\delta_d^{} \,\,=\,\, \frac{M_{12}^{d,\kappa}}{M_{12}^{d,\rm SM}} \,\,.
\end{eqnarray}
Accordingly, we impose  \,$-0.2\le{\rm Re}\,\delta_d^{}\le+0.02$,\,  which leads to
\begin{eqnarray} \label{bb1}
-0.031\,\,\le\,\,\, {\rm Re}\bigl(\kappa_{cb}^{\rm L}+\kappa_{cd}^{\rm L}\bigr) \,+\,
0.4\,{\rm Im}\bigl(\kappa_{cb}^{\rm L}-\kappa_{cd}^{\rm L}\bigr) \,\,\le\,\, 0.003 \,\,.
\end{eqnarray}

An additional constraint can be extracted from the $\beta$ measurement in \,$B\to J/\psi K$.\,
The anomalous couplings enter $\beta^{\rm eff}$ via both the mixing and decay amplitudes.
Since the mixing parameters $p_{B_d}$ and $q_{B_d}$ are related to $M_{12}^d$ by
\,$q_{B_d}/p_{B_d}\simeq M_{12}^{d*}/\bigl|M_{12}^d\bigr|$\,~\cite{Nir:1992wi}, we have
\begin{eqnarray}
\frac{q_{B_d}^{}}{p_{B_d}^{}} \,\,\simeq\,\,
\sqrt{\frac{M_{12}^{d,\rm SM*}\bigl(1+\delta_d^*\bigr)}
           {M_{12}^{d,\rm SM}\bigl(1+\delta_d^{}\bigr)}}
\,\,\simeq\,\, \frac{V_{td}^{}V_{tb}^*}{V_{td}^*V_{tb}^{}}\, e^{-i\,{\rm Im}\,\delta_d^{}} \,\,.
\end{eqnarray}
From the decay amplitude in Eq.~(\ref{amp}),  we derive
\begin{eqnarray} \label{ampd}
\frac{{\cal M}\bigl(\bar B^0\to\psi K_S^{}\bigr)}{{\cal M}\bigl(B^0\to\psi K_S^{}\bigr)} \,\,=\,\,
-\frac{V_{cd}^*V_{cb}^{}\bigl(1+\kappa_{cs}^{\rm L*}+\kappa_{cb}^{\rm L}\bigr)}
      {V_{cd}^{}V_{cb}^*\bigl(1+\kappa_{cs}^{\rm L}+\kappa_{cb}^{\rm L*}\bigr)} \,\,\simeq\,\,
-\frac{V_{cd}^*V_{cb}^{}}{V_{cd}^{}V_{cb}^*}
\bigl[1+2i\,{\rm Im}\bigl(\kappa_{cb}^{\rm L}-\kappa_{cs}^{\rm L}\bigr)\bigr] \,\,,
\end{eqnarray}
having incorporated the $K$-mixing factor
\,$q_K^{}/p_K^{}=V_{cd}^*V_{cs}^{}/\bigl(V_{cd}^{}V_{cs}^*\bigr)$\,~\cite{Nir:1992wi}.
It follows that
\begin{eqnarray}
e^{-2i\beta_{\psi K}^{\rm eff}} \,\,=\,\,  \frac{q_{B_d}^{}}{p_{B_d}^{}}\,
\frac{{\cal M}\bigl(\bar B^0\to\psi K_S^{}\bigr)}{{\cal M}\bigl(B^0\to\psi K_S^{}\bigr)}
\,\,\simeq\,\,  e^{-2i\beta^{\rm SM}}\,
e^{2i\,{\rm Im}(\kappa_{cb}^{\rm L}-\kappa_{cs}^{\rm L})-i\,{\rm Im}\,\delta_d^{}} \,\,.
\end{eqnarray}
Upon comparing the experimental value \,$2\beta^{\rm eff}=2\beta_{\psi K}^{\rm eff}=0.717\pm0.033$\,
from Eq.~(\ref{betad}) to the SM prediction
\,$2\beta^{\rm SM}=0.753_{-0.028}^{+0.032}$\,~\cite{ckmfit}, we then require
\,$-0.01\le2\,{\rm Im}(\kappa_{cb}^{\rm L}-\kappa_{cs}^{\rm L})-{\rm Im}\,\delta_d^{}\le0.08$,\,
which implies
\begin{eqnarray} \label{2b1}
-1.5\times10^{-3} \,\,\le\,\, 0.4\, {\rm Re}\bigl(\kappa_{cb}^{\rm L}+\kappa_{cd}^{\rm L}\bigr)
- 0.69\, {\rm Im}\,\kappa_{cb}^{\rm L} + {\rm Im}\,\kappa_{cd}^{\rm L}
- 0.31\, {\rm Im}\,\kappa_{cs}^{\rm L} \,\,\le\,\, 0.012 \,\,.
\end{eqnarray}

\subsection{\boldmath$B_s$-$\bar B_s$ mixing}

The SM part of the matrix element $M_{12}^s$ for $B_s^0$-$\bar B_s^0$ mixing is also dominated
by the top contribution~\cite{Buchalla:1995vs},
\begin{eqnarray}
M_{12}^{s,\rm SM} \,\,\simeq\,\,
\frac{G_{\rm F}^2\,m_W^2}{12\pi^2}\, f_{B_s}^2 m_{B_s}^{}\, \eta_B^{}B_{B_s}^{}\,
\bigl(V_{tb}^{}V_{ts}^*\bigr)^2\,S_0^{}\bigl(x_t^{}\bigr) \,\,.
\end{eqnarray}
For the anomalous couplings, again the charm contribution alone is
\begin{eqnarray}
M_{12}^{s,\kappa} &=&
\frac{G_{\rm F}^2 m_W^2}{24\pi^2}\, f_{B_s}^2 m_{B_s}^{}\, \eta_B^{}B_{B_s}^{}\,
\lambda_c^{sb}\bigl(\kappa_{cb}^{\rm L}+\kappa_{cs}^{\rm L*}\bigr)
\Biggl( -\lambda_t^{sb}\, x_t^{}\,\ln\frac{\Lambda^2}{m_W^2} -
\sum_q\lambda_q^{sb}\, {\cal B}_1^{}\bigl(x_q^{},x_c^{}\bigr) \Biggr)
\nonumber \\ &=&
(53-0.95\,i)\,{\rm ps}^{-1}\, \bigl(\kappa_{cb}^{\rm L}+\kappa_{cs}^{\rm L*}\bigr) \,\,,
\end{eqnarray}
where  \,$\lambda_q^{sb}=V_{qs}^*V_{qb}^{}$.\,

Similarly to the $B_d^{}$ case, we have here
\begin{eqnarray} \label{ds}
\Delta M_s^{\rm exp}\,\,=\,\, \Delta M_s^{\rm SM}\,\bigl|1+\delta_s^{}\bigr| \,\,, \hspace{5ex}
\delta_s^{} \,\,=\,\, \frac{M_{12}^{s,\kappa}}{M_{12}^{s,\rm SM}} \,\,.
\end{eqnarray}
The experimental value, \,$\Delta M_s^{\rm exp}=(17.77\pm0.12)\,{\rm ps}^{-1}$\,~\cite{pdg},
is in agreement with the SM prediction,
\,$\Delta M_s^{\rm SM}=\bigl(17.6^{+1.7}_{-1.8}\bigr)\,{\rm ps}^{-1}$\,~\cite{ckmfit}.
These numbers allow us to require  \,$-0.09\le{\rm Re}\,\delta_s^{}\le 0.1$,\,  leading to
\begin{eqnarray} \label{bb2}
-0.014\,\,\le\,\,{\rm Re}\bigl(\kappa_{cs}^{\rm L}+\kappa_{cb}^{\rm L}\bigr) \,+\,
0.018\,{\rm Im}\bigl(\kappa_{cs}^{\rm L}-\kappa_{cb}^{\rm L}\bigr) \,\,\le\,\, 0.015 \,\,.
\end{eqnarray}

A complementary constraint is provided by the parameter $\beta_s^{}$ in $B_s^{}$ decay,
analogously to $\beta$ in $B_d^{}$ decay.
In this case, the mode of interest is \,$\bar B_s^0\to J/\psi\phi$,\, which proceeds from the same
\,$b\to sc\bar c$\, transition as  \,$\bar B_d^0\to J/\psi\bar K$.\,
For the mixing factor, we have
\begin{eqnarray}
\frac{q_{B_s}^{}}{p_{B_s}^{}} \,\,\simeq\,\,
\sqrt{\frac{M_{12}^{s,\rm SM*}\bigl(1+\delta_s^*\bigr)}
           {M_{12}^{s,\rm SM}\bigl(1+\delta_s^{}\bigr)}} \,\,\simeq\,\,
\frac{V_{ts}^{}V_{tb}^*}{V_{ts}^*V_{tb}^{}}\, e^{-i\,{\rm Im}\,\delta_s^{}} \,\,,
\end{eqnarray}
and for the ratio of decay amplitudes, as in Eq.~(\ref{ampd}),
\begin{eqnarray}
\frac{{\cal M}\bigl(\bar B_s^0\to(J/\psi\phi)_f^{}\bigr)}
     {{\cal M}\bigl(B_s^0\to(J/\psi\phi)_f^{}\bigr)}
\,\,\simeq\,\, \eta_f^{}\, \frac{V_{cs}^*V_{cb}^{}}{V_{cs}^{}V_{cb}^*}
\bigl[1+2i\,{\rm Im}\bigl(\kappa_{cb}^{\rm L}-\kappa_{cs}^{\rm L}\bigr)\bigr] \,\,,
\end{eqnarray}
where  $(J/\psi\phi)_f$  is one of the $CP$ eigenstates of the $J/\psi\phi$ final-state
and $\eta_f^{}$  its $CP$ eigenvalue.
It follows that
\begin{eqnarray}
e^{2i\beta_{\psi\phi}^{\rm eff}} \,\,=\,\,  -\eta_f^{}\, \frac{q_{B_s}^{}}{p_{B_s}^{}}\,
\frac{{\cal M}\bigl(\bar B_s^0\to(J/\psi\phi)_f^{}\bigr)}
     {{\cal M}\bigl(B_s^0\to(J/\psi\phi)_f^{}\bigr)}
\,\,\simeq\,\,  e^{2i\beta_s^{\rm SM}}\,
e^{2i\,{\rm Im}(\kappa_{cb}^{\rm L}-\kappa_{cs}^{\rm L})-i\,{\rm Im}\,\delta_s^{}} \,\,.
\end{eqnarray}
The SM yields  \,$2\beta_s^{\rm SM}=0.03614^{+0.00172}_{-0.00162}$\,~\cite{ckmfit}, but
the measurements of  \,$B_s^{}\to J/\psi\phi$\, yield the average value
\,$2\beta_s^{\rm eff}=2\beta_{\psi\phi}^{\rm eff}=0.77_{-0.29}^{+0.37}$ or
$2.36_{-0.37}^{+0.29}$\,~\cite{hfag}.
It is again too early to attribute this difference to new physics, but it can be used to impose
the bound
\,$-0.003\le 2\,{\rm Im}(\kappa_{cb}^{\rm L}-\kappa_{cs}^{\rm L})-{\rm Im}\,\delta_s^{}\le 0.4$,\,
which yields
\begin{eqnarray} \label{2b2}
-0.09 \,\,\le\,\, 0.026\, {\rm Re}\bigl(\kappa_{cb}^{\rm L}+\kappa_{cs}^{\rm L}\bigr) \,+\,
{\rm Im}\bigl(\kappa_{cb}^{\rm L}-\kappa_{cs}^{\rm L}\bigr) \,\,\le \,\, 7\times10^{-4} \,\,.
\end{eqnarray}

\section{Summary and conclusions\label{final}}

We have explored the phenomenological consequences of anomalous $W$-boson couplings to the charm
quark in a comprehensive way.
Most of the constraints we have obtained are summarized in Table~\ref{t:summary}.
In writing them, we have followed the discussion in Appendix~\ref{Keff} about the independent
parameters in the quark-mixing matrices and chosen \,$\arg\kappa_{cd}^{\rm L}=0$.\,
Consequently, we have used  the condition \,${\rm Im}\,\kappa_{cd}^{\rm L}=0$\, in all the results.
All the constraints in this table are quoted as 1-$\sigma$ errors, but in some cases the theoretical
error is only an order of magnitude and this is not reflected in the quoted range.
The discussion in the text makes it clear whenever this happens.
We leave out from the table the processes  \,$b\to s\gamma$\, and \,$s\to d \gamma$\,  since
the resulting bounds are not competitive with the rest.

In Fig.~\ref{kcs,kcb} we show the parameter space of the real and imaginary parts of
$\kappa_{cs}^{\rm L}$ and $\kappa_{cb}^{\rm L}$ assuming that only one of them is nonzero at a time.
This figure indicates that currently the phase of $\kappa_{cs}^{\rm L}$ is only loosely
constrained and ranges from $-90^\circ$ to $90^\circ$.
In contrast, the phase of $\kappa_{cb}^{\rm L}$ is unconstrained if its magnitude is small (at
the $10^{-3}$ level).  However, larger values of $|\kappa_{cb}^{\rm L}|$, at the few percent level,
are also allowed provided its phase lies in a range roughly between $-150^\circ$ and~$-56^\circ$.

\begin{table}
\centering \footnotesize \renewcommand{\arraystretch}{1.4}
\begin{tabular}{|c|c|c|c|}
\hline\hline
Process & Eq. & Constraint & \# \\
\hline
$D\to\ell\nu$ &  \vphantom{$\big|^{\big|}$} (\ref{recd}) &
$\bigl|{\rm Re}\bigl(\kappa_{cd}^{\rm L}-\kappa_{cd}^{\rm R}\bigr)\bigr| \le 0.04$ & 1 \\
$D_s\to\ell\nu$ & (\ref{recs}) &
$0 \le {\rm Re}\bigl(\kappa_{cs}^{\rm L}-\kappa_{cs}^{\rm R}\bigr) \le 0.1$ & 2 \\
$b\to c\ell\bar\nu$ & (\ref{recb}) &
$-0.13 \le {\rm Re}\,\kappa^{\rm R}_{cb} \le 0$ & 3 \\
$B\to J/\psi K, \eta_c^{} K$ & (\ref{im(cb+cs)}) &
$-5\times10^{-4} \le {\rm Im}\bigl(\kappa_{cb}^{\rm R}+\kappa_{cs}^{\rm R}\bigr) \le 0.04$ & 4 \\
$K^+\to \pi^+ \nu\bar{\nu}$ & (\ref{k2pnn}) &
$-1.3\times10^{-3} \le {\rm Re}\bigl(\kappa^{\rm L}_{cd}+\kappa^{\rm L}_{cs}\bigr) +
0.42\,{\rm Im}\,\kappa^{\rm L}_{cs} \le 2.5\times10^{-4}$ & 5 \\
$K_L \to \mu^+ \mu^-$ & (\ref{k2ll}) &
$\Bigl|{\rm Re}\bigl(\kappa_{cs}^{\rm L}+\kappa_{cd}^{\rm L}\bigr) +
6\times10^{-4}\,{\rm Im}\,\kappa_{cs}^{\rm L}\Bigr|
\le 1.5\times 10^{-4}$ & 6 \\
$\Delta M_K$ & (\ref{kk1}) &
$\bigl|0.043\, {\rm Re}\bigl(\kappa_{cd}^{\rm L}+\kappa_{cs}^{\rm L}\bigr)
- 0.015\,{\rm Im}\,\kappa_{cs}^{\rm L}
- {\rm Re}\bigl(\kappa_{cd}^{\rm R*}\kappa_{cs}^{\rm R}\bigr)
+ 0.28\,{\rm Im}\bigl(\kappa_{cd}^{\rm R*}\kappa_{cs}^{\rm R}\bigr) \bigr| \le 8.5\times10^{-4}$ & 7 \\
$\epsilon$ (mixing) & (\ref{kk2}) &
\,$\bigl|0.015\, {\rm Re}\bigl(\kappa_{cs}^{\rm L}+\kappa_{cd}^{\rm L}\bigr)
+ 0.043\,{\rm Im}\,\kappa_{cs}^{\rm L}
- 0.28\,{\rm Re}\bigl(\kappa_{cd}^{\rm R*}\kappa_{cs}^{\rm R}\bigr)
- {\rm Im}\bigl(\kappa_{cd}^{\rm R*}\kappa_{cs}^{\rm R}\bigr) \bigr| \le 2.5\times10^{-6}$\, & 8 \\
$\Delta M_d$ & (\ref{bb1}) &
$-0.031\le {\rm Re}\bigl(\kappa_{cb}^{\rm L}+\kappa_{cd}^{\rm L}\bigr) +
0.4\,{\rm Im}\,\kappa_{cb}^{\rm L} \le 0.003 $ & 9 \\
$\sin(2\beta)$ (mixing) & (\ref{2b1}) &
$-1.5\times10^{-3} \le 0.4\, {\rm Re}\bigl(\kappa_{cb}^{\rm L}+\kappa_{cd}^{\rm L}\bigr)
- 0.69\, {\rm Im}\,\kappa_{cb}^{\rm L} - 0.31\, {\rm Im}\,\kappa_{cs}^{\rm L} \le 0.012$ & 10 \\
$\Delta M_s$ & (\ref{bb2}) &
$-0.014\le{\rm Re}\bigl(\kappa_{cs}^{\rm L}+\kappa_{cb}^{\rm L}\bigr) +
0.018\,{\rm Im}\bigl(\kappa_{cs}^{\rm L}-\kappa_{cb}^{\rm L}\bigr) \le 0.015$ & 11 \\
\,$\sin(2\beta_s)$ (mixing)\, & \,(\ref{2b2})\, &
$-0.09 \le 0.026\, {\rm Re}\bigl(\kappa_{cb}^{\rm L}+\kappa_{cs}^{\rm L}\bigr) +
{\rm Im}\bigl(\kappa_{cb}^{\rm L}-\kappa_{cs}^{\rm L}\bigr) \le 7\times10^{-4} $ & \,12\, \\ [0.5ex]
\hline\hline
\end{tabular}
\caption{Summary of constraints, with their equation numbers, from various processes.}
\label{t:summary}
\end{table}

\begin{figure}[ht]
\includegraphics[width=3.3in]{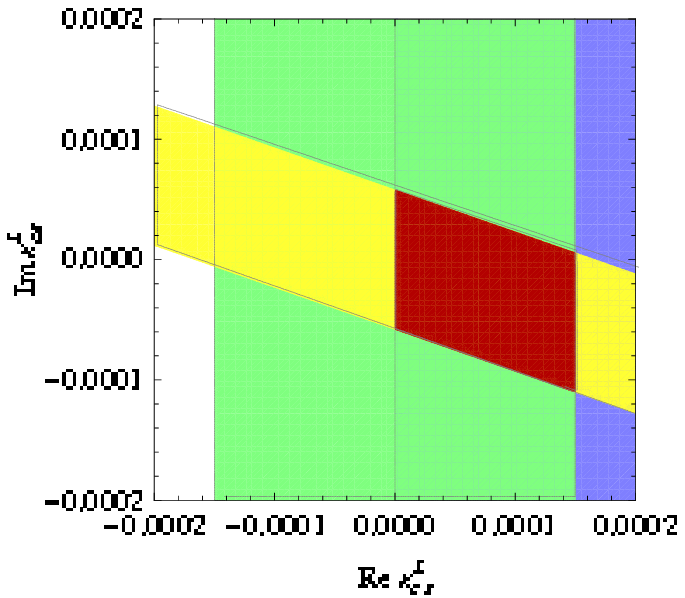}
\includegraphics[width=3.1in]{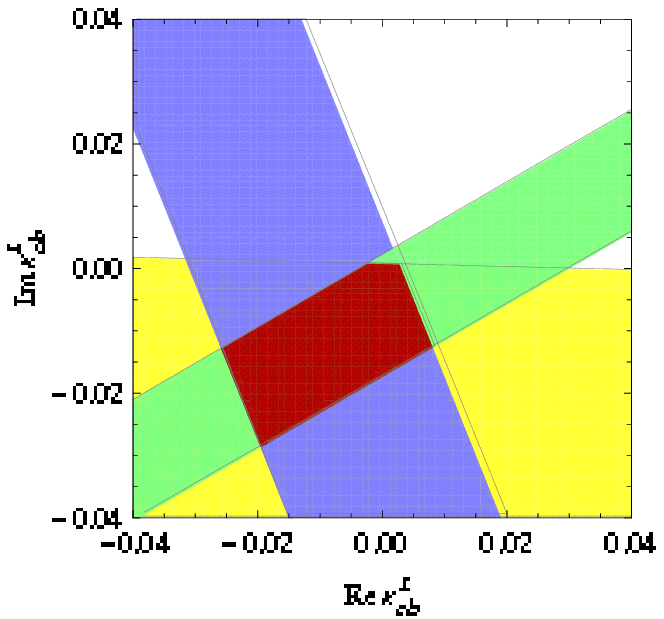}
\caption{\label{kcs,kcb}
Parameter space of the real and imaginary parts of $\kappa_{cs}^{\rm L}$ and $\kappa_{cb}^{\rm L}$
subject to the relevant constraints in Table~\ref{t:summary},
under the assumption that only one $\kappa$ is nonzero at at time.
The heavily (blue), medium (green), and lightly (yellow) shaded areas in the left plot satisfy
constraints {\scriptsize\#}2, {\scriptsize\#}6, and {\scriptsize\#}8, respectively.
The heavily (blue), medium (green), and lightly (yellow) shaded in the right plot satisfy
constraints {\scriptsize\#}9, {\scriptsize\#}10, and {\scriptsize\#}12, respectively.
The dark (red) region in each plot satisfies all the constraints in it.}
\end{figure}

\begin{figure}[ht]
\includegraphics[width=3.31in]{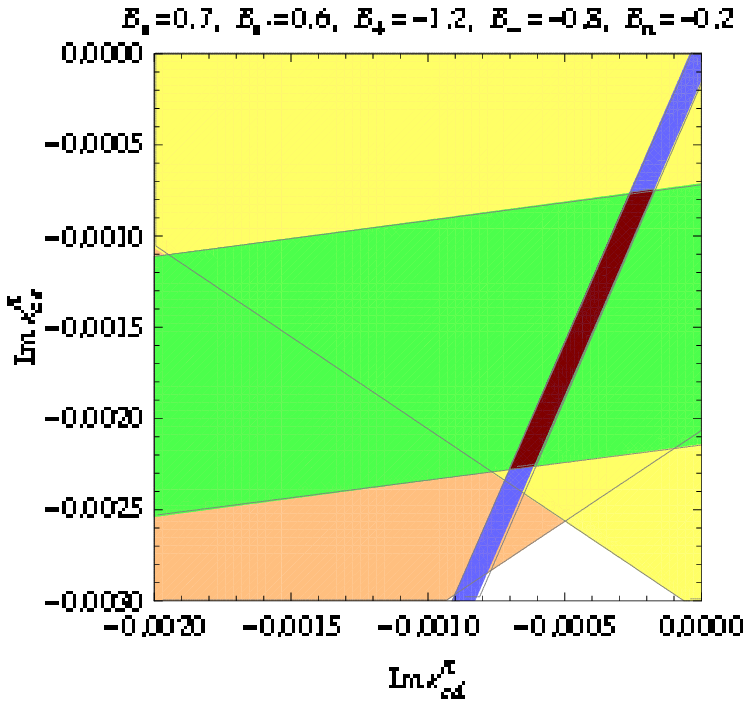}
\includegraphics[width=3.23in]{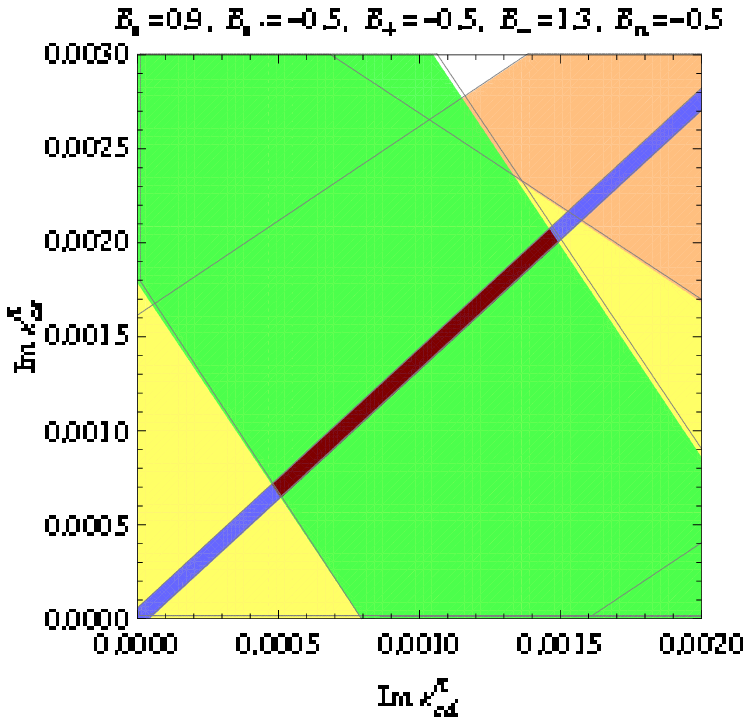}
\caption{\label{kplots}
Parameter space of ${\rm Im}\,\kappa_{cd}^{\rm R}$ and ${\rm Im}\,\kappa_{cs}^{\rm R}$ subject to
constraints from the contributions of magnetic-dipole operators to $\epsilon$, $\epsilon'$,
$A_{\Xi\Lambda}$, and the neutron EDM for two representative sets of $B_{\epsilon,\epsilon',+,-,n}$.
The various regions are described in the text.}
\end{figure}

We treat the constraints arising from the contributions of magnetic-dipole operators to
$CP$-violating observables separately and display those in Fig.~\ref{kplots}.
These observables receive contributions from the anomalous couplings
${\rm Im}\,\kappa_{cd,cs}^{\rm R}$ that are much larger than SM contributions to
the dipole operators. These enhanced contributions to $\epsilon$, $\epsilon'$,  the neutron EDM,
and $A_{\Xi\Lambda}^{}$ have been studied before as they arise within LR models and
supersymmetry~\cite{He:1999bv,Buras:1999da,Chang:1994wk}.

The calculations for all of these $CP$-violating observables suffer from large theoretical
uncertainties which we have parametrized with $B$ factors in this paper.
For illustration, we display two plots in Fig.~\ref{kplots} resulting from choosing two
representative sample sets of values of the parameters $B_{\epsilon,\epsilon',+,-,n}$ within
their ranges in Eqs.~(\ref{bebep}), (\ref{b+b-}), and~(\ref{bn}) and imposing the constraints
in Eqs.~(\ref{eep}), (\ref{ax}), and~(\ref{dn_bound}).
In each of the plots, the very lightly shaded (yellow) band satisfies the $\epsilon'/\epsilon$
constraint, the lightly shaded (pink) band the $\epsilon$ constraint, the medium shaded (green)
band the $A_{\Xi\Lambda}^{}$ constraint, the heavily shaded (blue) band the $d_n^{}$ constraint,
and the dark (red) region all of the constraints.
It is worth noting that there is a significant amount of the parameter space where all of
the constraints can be simultaneously satisfied and that the values of
${\rm Im}\,\kappa_{cd,cs}^{\rm R}$ involved are typically of order a few times $10^{-3}$ or less.
Furthermore, as is obvious from the plots, the neutron-EDM constraint is the most restrictive.
Also, interestingly, the allowed region of parameter space easily accommodates
an $A_{\Xi\Lambda}^{}$ much larger than the SM prediction, as hinted at by the preliminary
measurement by HyperCP~\cite{hypercp}.

In order to gain some insight into the constraints in Table~\ref{t:summary} and Fig.~\ref{kplots},
we have extracted the ranges corresponding to taking only one anomalous coupling at a time to be
nonzero (and only for the cases of a purely real or a purely imaginary coupling).
They are collected in Table~\ref{t:oneatatime}.
This table shows that, in general, the left-handed couplings are much more constrained than
the right-handed couplings. Similarly, the imaginary part of the couplings is more tightly
constrained than the corresponding real part. The largest deviations allowed by current data
appear in the real part of the right-handed couplings, which can be as large as 10\% of
the corresponding SM couplings.

\begin{table}[t]
\medskip \centering \renewcommand{\arraystretch}{1.5}
\begin{tabular}{|c|c||c|c|}
\hline\hline
$0 \le {\rm Re}\,\kappa_{cd}^{\rm L} \le 1.5\times 10^{-4}$ \, & {\scriptsize\#}6 &
$\bigl({\rm Im}\,\kappa_{cd}^{\rm L}=0\bigr)$ & --- \\
$0 \le{\rm Re}\,\kappa_{cs}^{\rm L}\le 1.5\times 10^{-4}$& {\scriptsize\#}2, {\scriptsize\#}6 &
$-6\times 10^{-5} \le {\rm Im}\,\kappa_{cs}^{\rm L} \le 6\times 10^{-5}$& {\scriptsize\#}8 \\
\, $-4\times 10^{-3}\le {\rm Re}\,\kappa_{cb}^{\rm L}\le 3\times 10^{-3}$ \, & \,
{\scriptsize\#}9, {\scriptsize\#}10 \, &
$-0.02\le{\rm Im}\,\kappa_{cb}^{\rm L}\le 7\times10^{-4}$&  {\scriptsize\#}10, {\scriptsize\#}12  \\
$-0.04 \le {\rm Re}\,\kappa_{cd}^{\rm R}\le 0.04 $& {\scriptsize\#}1 &
\, $-2\times 10^{-3}\le {\rm Im}\,\kappa_{cd}^{\rm R}\le2\times 10^{-3}$ \, & Fig.~\ref{kplots}\\
$-0.1\le {\rm Re}\,\kappa_{cs}^{\rm R}\le 0$& {\scriptsize\#}2 &
$-5\times10^{-4}\le{\rm Im}\,\kappa_{cs}^{\rm R}\le2\times 10^{-3}$& \, {\scriptsize\#}4, Fig.~\ref{kplots} \, \\
$-0.13\le {\rm Re}\,\kappa_{cb}^{\rm R}\le 0$& {\scriptsize\#}3 &
$-5\times10^{-4}\le{\rm Im}\,\kappa_{cb}^{\rm R}\le 0.04$& {\scriptsize\#}4 \\
\hline\hline
\end{tabular}
\caption{Constraints on each of the anomalous charm couplings, extracted from
Table~\ref{t:summary} and Fig.~\ref{kplots}.}
\label{t:oneatatime}
\end{table}

For specific model building, it is useful to recall that the new physics in the quark-mixing
matrices is being parametrized here as the product of the anomalous coupling and the corresponding
CKM matrix element, as can be seen in Eqs.~(\ref{effmatrix}) and~(\ref{effmatr}).
The allowed new physics in the left-handed sector is at most of ${\cal O}\bigl(\lambda^4\bigr)$
for the $cb$ coupling and of ${\cal O}\bigl(\lambda^6\bigr)$ for the other two,
\,$\lambda\sim0.23$\,  being the usual Cabibbo parameter.
This conclusion also implies that current data allow deviations from unitarity in the quark-mixing
matrix only at ${\cal O}\bigl(\lambda^5\bigr)$ or higher.
On the other hand, new physics affecting right-handed quarks can be of
${\cal O}\bigl(\lambda^3\bigr)$ for $cd$ and $cb$ transitions, and as large as
${\cal O}(\lambda)$ for $cs$ transitions.  That is to say that the right-handed $cs$
matrix-element is the least constrained.

Perhaps surprisingly, we note that the constraints displayed in Table~\ref{t:oneatatime} are
comparable or tighter than existing constraints on anomalous $W$-boson couplings to the top
quark~\cite{Fujikawa:1993zu}.
We can gain more insight into these numbers by interpreting them in the context of left-right (LR)
models with mixing of the $W_{\rm L}$ and $W_{\rm R}$ gauge bosons.
In these models one predicts  \,$\kappa^{\rm L}\simeq -\frac{1}{2}\xi_W^2$,\, where
$\xi_W^{}$ is the $W_{\rm L}$-$W_{\rm R}$ mixing angle.
This angle is constrained by \,$b\to s\gamma$\, to be at the $10^{-3}$ level~\cite{Langacker:1989xa},
and so $\kappa^{\rm L}$ in these models is only allowed at the $10^{-6}$ level.
The additional freedom found in our study arises from the general decoupling between the top
and charm anomalous couplings.

For the right-handed anomalous charm couplings, the LR models result in the generic form
\,$\kappa^{\rm R}_{cD}\simeq\bigl(g_{\rm R}^{}/g_{\rm L}^{}\bigr)\xi_W^{}V_{cD}^{\rm R}/V_{cD}^{}$\,
for \,$D=d,s,b$.\,
The first factor, $\bigl(g_{\rm R}^{}/g_{\rm L}^{}\bigr)\xi_W^{}$, is allowed to be several times
larger than $\xi_W^{}$~\cite{Langacker:1989xa}, whereas the second
factor depends on the right-handed mixing matrix,~$V^{\rm R}$.
Our bounds in Table~\ref{t:oneatatime} suggest within this context that constraints on right-handed
mixing-matrix elements involving charm, $V_{cD}^{\rm R}$, are not very tight at present,
with $V_{cs}^{\rm R}$ being the least constrained~one.

Finally, our study also indicates which future measurements provide the most sensitive tests for
new physics that can be parametrized with anomalous charm-$W$ couplings.
For the $CP$-violating imaginary parts, the $n$ EDM and the hyperon asymmetry $A_{\Xi\Lambda}^{}$
are the most promising channels for probing right-handed couplings.
To probe $CP$-violating left-handed couplings, more precise measurements of $\sin(2\beta)$ and
$\sin(2\beta_s)$ are desired.
Constraints on the real parts of the right-handed couplings can be tightened with improved
measurements of semileptonic $B$ and $D$ decays.

\acknowledgments \vspace*{-1ex} 

The work of X.G.H. and J.T. was supported in part by NSC and NCTS.
The work of G.V. was supported in part by DOE under Contract No.~DE-FG02-01ER41155.
We thank David Atwood for useful conversations.

\appendix

\section{Independent parameters in quark mixing matrices\label{Keff}}

The parametrization we introduced in Eq.~(\ref{ludw}) reflects two effective matrices,
$K^{\rm L,R}_{\rm eff}$,  for the charged currents involving left- and right-handed quarks,
respectively.  In the effective theory under consideration, these 3$\times$3 matrices are complex
and nonunitary; the usual unitarity of the CKM matrix is lost.
The general 3$\times$3 complex matrix has 18 parameters: 9 magnitudes and 9 phases.
Nevertheless, not all the parameters in the matrix describing the left-handed charged current
are independent.  Of the 9 phases, 5 can be removed by redefinitions of the quark fields.
Considering a scenario in which the corrections to the CKM picture in the SM are small, we find it
convenient to choose these parameters as
\begin{eqnarray}
K^{\rm L}_{\rm eff} \,\,= \left( \begin{array}{ccc}
V_{ud}^{} & V_{us}^{} & V_{ub}^{}\,e^{i\phi_{ub}} \vspace{1ex} \\
V_{cd}^{}\Bigl(1+\bigl|\kappa_{cd}^{\rm L}\bigr|\,e^{i\phi_{cd}^{\rm L}}\Bigr) \, & \,
V_{cs}^{}\Bigl(1+\bigl|\kappa_{cs}^{\rm L}\bigr|\,e^{i\phi_{cs}^{\rm L}}\Bigr) \, & \,
V_{cb}^{}\Bigl(1+\bigl|\kappa_{cb}^{\rm L}\bigr|\,e^{i\phi_{cb}^{\rm L}}\Bigr) \vspace{1ex} \\
V_{td}^{}\,e^{i\phi_{td}} & V_{ts}^{} & V_{tb}^{}
\end{array} \right) .
\end{eqnarray}
More explicitly, four of the five quark phases have been used to remove the phases in
$V_{ud}$, $V_{us}$, $V_{ts}$, and $V_{tb}$.
The remaining quark phase has to be used to remove one of the phases in the charm row, and for
convenience we choose it to be $\phi_{cd}^{}$, thus setting it to zero. Equivalently, only the two
relative phases between the three $\bigl(K^{\rm L}_{\rm eff}\bigr){}_{ci}^{}$ elements are physical.
These, plus the phases $\phi_{ub}^{}$ and $\phi_{td}^{}$ of $V_{ub}$ and $V_{td}$ which become
independent, are the four physical phases in the general (nonunitary) 3$\times$3 complex matrix.
In this study we only  allow for three physical phases by requiring that the phases of $V_{ub}$
and $V_{td}$ be related, as in the usual CKM picture.

There are also nine independent real parameters, the magnitudes of each of the elements of
$K^{\rm L}_{\rm eff}$.  For our study we concentrate on the possibility of new physics in the charm
sector, leading us to reduce the number of free parameters that we consider from 9 to 6 by simply
setting the other 3 to zero.  We choose 3 of the 6 to be the 3 angles that describe the usual CKM
matrix and the rest the 3 anomalous couplings that appear explicitly in $K^{\rm L}_{\rm eff}$.
We are thus left with
\begin{eqnarray}
K^{\rm L}_{\rm eff}\,\,= \left( \begin{array}{ccc} \displaystyle
1-\frac{\lambda^2}{2} & \lambda & A\lambda^3(\rho -i \eta) \vspace{1ex} \\
-\lambda\bigl(1+\bigl|\kappa^{\rm L}_{cd}\bigr|\bigr) & \displaystyle
\biggl(1-\frac{\lambda^2}{2}\biggr)\Bigl(1+\bigl|\kappa^{\rm L}_{cs}\bigr|\,e^{i\phi^{\rm L}_{cs}}\Bigr)
\, & \,
A\lambda^2\Bigl(1+\bigl|\kappa^{\rm L}_{cb}\bigr|\,e^{i\phi^{\rm L}_{cb}}\Bigr) \vspace{1ex} \\
A\lambda^3(1-\rho-i\eta) \, & -A\lambda^2 & 1
\end{array} \right),
\label{effmatrix}
\end{eqnarray}
where, for simplicity of notation, we have included here only the usual terms
up to order $\lambda^3$ in the Wolfenstein parametrization of the CKM matrix;
the complete, unitary CKM matrix should be understood as remaining in Eq.~(\ref{effmatrix})
in the limit  \,$\kappa_{ci}^{}\to 0$\, and \,$\phi_{cj}^{}\to 0$.\,
In this paper we have provided formulas for the relevant observables in terms of these nine
parameters.
This set of formulas would allow us, in principle, to repeat the global fits for this
scenario and constrain the nine parameters.  We content ourselves with a less ambitious analysis
in which we assume, in accord with observation, that the CKM picture is dominant and deviations
are small.  We thus use the CKM parameters as extracted from the global fits of
CKMfitter~\cite{ckmfit} as input.  The subsequent comparison of this global fit with specific
observables then reflects the extent to which deviations from unitarity are allowed by current data.
Conservatively, we carry out this comparison at the one-sigma level.

None of the above considerations apply to the right-handed charged current: all the SM field phases
are fixed by the removal of five phases in $K^{\rm L}_{\rm eff}$ so that all 18 parameters in
the corresponding $K^{\rm R}_{\rm eff}$ are physical.
In this study we limit ourselves to 6 of them, which we write as
\begin{eqnarray}
K^{\rm R}_{\rm eff} \,\,= \left( \begin{array}{ccc}
0 & 0 & 0 \vspace{1ex} \\ \displaystyle
-\lambda\, \bigl|\kappa^{\rm R}_{cd}\bigr|\,e^{i\phi^{\rm R}_{cd}} \, & \, \displaystyle
\biggl(1-\frac{\lambda^2}{2}\biggr)\, \bigl|\kappa^{\rm R}_{cs}\bigr|\,e^{i\phi^{\rm R}_{cs}} \, & \,
A\lambda^2\, \bigl|\kappa^{\rm R}_{cb}\bigr|\,e^{i\phi^{\rm R}_{cb}} \vspace{1ex} \\
0 & 0 & 0
\end{array} \right).
\label{effmatr}
\end{eqnarray}
As pointed out in the conclusion, one needs to keep in mind with this parametrization that
the new physics is not just $\kappa_{ij}^{}$ but its product with $V_{ij}^{}$, as explicitly seen
in Eq.~(\ref{effmatr}).

\section{Amplitudes in unitary gauge\label{unitary}}

\begin{figure}[bh]
\includegraphics{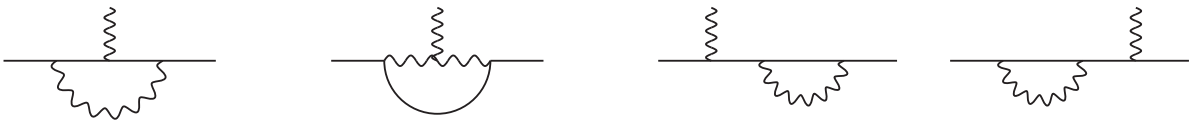}
\caption{\label{penguins}Diagrams contributing to amplitudes for \,$d\to d'{\cal V}^*$,\, with
$\cal V$ being a neutral gauge boson. In all diagram figures, straight lines denote fermions and
the loops contain $W$ bosons besides fermions.}
\end{figure}

\begin{figure}[hb]
\includegraphics{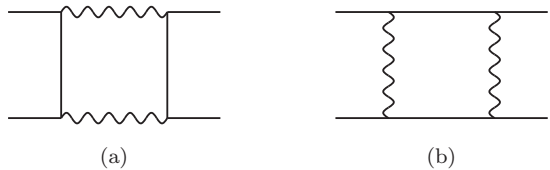}
\caption{\label{boxes}Box diagrams contributing to amplitudes for (a) $d\bar d'\to \ell^+\ell^-$ or
$\bar\nu\nu$\, and (a,b) $d\bar d'\to d\bar d'$.}
\end{figure}

The loop diagrams that are relevant to some of the processes we consider are displayed in
Figs.~\ref{penguins} and~\ref{boxes}.  In evaluating the diagrams, we use dimensional
regularization with a completely anticommuting~$\gamma_5^{}$ and adopt the unitary gauge, which
implies that they contain only fermions and $W$-bosons.
Since the theory with anomalous couplings is not renormalizable, some of the one-loop results
are divergent.
For these we adopt the prescription to drop the combination \,$2/(4-D)-\gamma_{\rm E}^{}+\ln(4\pi)$\,
in the $D$-dimensional integral, $\gamma_{\rm E}^{}$ being the Euler constant,
and retain the accompanying logarithmic, $\ln\bigl(\mu/m_W^{}\bigr)$, part as well as other
finite terms that depend on the mass of the quark in the loop.
Moreover, we identify the renormalization scale $\mu$ with  the scale $\Lambda$ of the new
physics parametrized by the anomalous
couplings, \,$\mu=\Lambda$.\,
In the SM limit $\bigl(\kappa^{\rm L,R}=0\bigr)$, after the unitarity relation
\,$V_{ud'}^*V_{ud}^{}+V_{cd'}^*V_{cd}^{}+V_{td'}^*V_{td}^{}=0$\, is imposed, our results are finite
and reproduce those obtained in the literature in $R_\xi$ gauges~\cite{Inami:1980fz}.


We will present the details of our calculation elsewhere~\cite{paper2}.
Here we provide the resulting effective Hamiltonians relevant to the loop-induced
processes dealt with in this paper.

The effective Hamiltonians for electromagnetic and chromomagnetic dipole operators involving
down-type quarks~$d$ and \,$d'\neq d$\, are derived from the four diagrams in Fig.~\ref{penguins}
with up-type quarks~$q$ in the loops and can be expressed as
\begin{eqnarray} \label{Hdd'f}
{\cal H}_{d\to d'\gamma}^{} \,\,=\,\,
\frac{-e G_{\rm F}^{}}{4\sqrt2\,\pi^2}\sum_{q=u,c,t} \bar d'\sigma_{\mu\nu}^{}
\bigl(F_{\rm L}^q P_{\rm L}^{}+F_{\rm R}^q P_{\rm R}^{}\bigr)d\,F^{\mu\nu} \,\,,
\end{eqnarray}
\begin{eqnarray} \label{Hdd'g}
{\cal H}_{d\to d'g}^{} \,\,=\,\,
\frac{-g_{\rm s}^{}G_{\rm F}^{}}{4\sqrt2\,\pi^2}\sum_{q=u,c,t} \bar d'\sigma_{\mu\nu}^{}
\bigl(G_{\rm L}^q P_{\rm L}^{}+G_{\rm R}^q P_{\rm R}^{}\bigr)t_a^{}d\,G_a^{\mu\nu} \,\,,
\end{eqnarray}
where  \,$\sigma^{\mu\nu}=\frac{i}{2}[\gamma^\mu,\gamma^\nu]$,\,
\begin{eqnarray} \label{FLR}
\begin{array}{c}   \displaystyle
F_{\rm L}^q \,\,=\,\,
\bigl(\lambda_q^{\rm L}\,m_{d'}^{}+\lambda_q^{\rm R}\,m_d^{}\bigr)F_0^{\rm SM}\big(x_q^{}\bigr) \,+\,
\lambda_q^{} \bigl(1+\kappa_{qd}^{\rm L}\bigr)\kappa_{qd'}^{\rm R*}\, m_q^{}\,
F_0^{}\big(x_q^{}\bigr) \,\,,
\vspace{2ex} \\   \displaystyle
F_{\rm R}^q \,\,=\,\,
\bigl(\lambda_q^{\rm L}\,m_d^{}+\lambda_q^{\rm R}\,m_{d'}^{}\bigr)F_0^{\rm SM}\big(x_q^{}\bigr) \,+\,
\lambda_q^{} \bigl(1+\kappa_{qd'}^{\rm L*}\bigr)\kappa_{qd}^{\rm R}\, m_q^{}\,
F_0^{}\big(x_q^{}\bigr) \,\,,
\end{array}
\end{eqnarray}
\begin{eqnarray} \label{GLR}
\begin{array}{c}   \displaystyle
G_{\rm L}^q \,\,=\,\,
\bigl(\lambda_q^{\rm L}\,m_{d'}^{}+\lambda_q^{\rm R}\,m_d^{}\bigr)G_0^{\rm SM}\big(x_q^{}\bigr) \,+\,
\lambda_q^{}\bigl(1+\kappa_{qd}^{\rm L}\bigr)\kappa_{qd'}^{\rm R*}\,m_q^{}\,G_0^{}\big(x_q^{}\bigr) \,\,,
\vspace{2ex} \\   \displaystyle
G_{\rm R}^q \,\,=\,\,
\bigl(\lambda_q^{\rm L}\,m_d^{}+\lambda_q^{\rm R}\,m_{d'}^{}\bigr)G_0^{\rm SM}\big(x_q^{}\bigr) \,+\,
\lambda_q^{}\bigl(1+\kappa_{qd'}^{\rm L*}\bigr)\kappa^{\rm R}_{qd}\,m_q^{}\,G_0^{}\big(x_q^{}\bigr) \,\,,
\end{array}
\end{eqnarray}
with
\begin{eqnarray}
\lambda_q^{} \,\,=\,\, V_{qd'}^*V_{qd}^{} \,\,, \hspace{3ex}
\lambda_q^{\rm L} \,\,=\,\, \lambda_q^{}\,
\bigl(1+\kappa_{qd'}^{\rm L*}\bigr)\bigl(1+\kappa_{qd}^{\rm L}\bigr) \,\,, \hspace{3ex}
\lambda_q^{\rm R}\,\,=\,\,\lambda_q^{}\,\kappa_{qd'}^{\rm R*}\kappa_{qd}^{\rm R} \,\,, \hspace{4ex}
x_f^{} \,\,=\,\, \frac{\bar m_f^2\bigl(m_f^{}\bigr)}{m_W^2} \,\,, \;\;\;
\end{eqnarray}
\begin{eqnarray} \label{f}
\begin{array}{c}   \displaystyle
F_0^{\rm SM}(x) \,\,=\,\, \frac{-7 x+5 x^2+8 x^3}{24(1-x)^3} \,-\, \frac{2 x^2-3 x^3}{4(1-x)^4}\,\ln x \,\,,
\vspace{2ex} \\   \displaystyle
F_0^{}(x) \,\,=\,\, \frac{-20+31 x-5 x^2}{12(1-x)^2} \,-\, \frac{2x-3x^2}{2(1-x)^3}\, \ln x \,\,,
\end{array}
\end{eqnarray}
\begin{eqnarray}  \label{g}
G_0^{\rm SM}(x) \,\,=\,\, \frac{-2 x-5 x^2+x^3}{8(1-x)^3} \,-\, \frac{3 x^2\,\ln x}{4(1-x)^4} \,\,,
\hspace{4ex}
G_0^{}(x) \,\,=\,\, \frac{-4-x-x^2}{4(1-x)^2} \,-\, \frac{3x\, \ln x}{2(1-x)^3}  \,\,.
\end{eqnarray}
We note that the loop calculation for these operators yields finite results.
We also note that different notations, \,$D_0'=-2F_0^{\rm SM}$,\, \,$E_0'=-2G_0^{\rm SM}$,\,
\,$\tilde F=F_0^{}$,\, and  \,$\tilde G=G_0^{}$,\,  are sometimes
used in the literature~\cite{Buchalla:1995vs,Cho:1993zb}.

The process \,$d\bar d'\to\nu\bar\nu$\, receives contributions from all the diagrams
in Fig.~\ref{penguins} via \,$d\bar d'\to Z^*\to\nu\bar\nu$\, and from the box diagram
in Fig.~\ref{boxes}(a), the loop fermions in the latter being an up-type quark and a lepton.
After summing over \,$q=u,c,t$\, and imposing the unitarity condition
\,$\lambda_u+\lambda_c+\lambda_t=0$,\,  we find the effective Hamiltonian
\begin{eqnarray}
{\cal H}_{d\bar d'\to\nu\bar\nu}^{} \,\,=\,\, {\cal H}_{d\bar d'\to\nu\bar\nu}^{\rm SM}
\,+\, {\cal H}_{d\bar d'\to\nu\bar\nu}^{\kappa} \,\,,
\end{eqnarray}
where
\begin{eqnarray} \label{dd2nn_sm}
{\cal H}_{d\bar d'\to\nu\bar\nu}^{\rm SM} \,\,=\,\,
\frac{\alpha\, G_{\rm F}^{}}{\sqrt8\, \pi\, \sin^2\theta_{\rm W}^{}}\sum_q
4 \lambda_q^{}\, X_0^{}\bigl(x_q^{}\bigr)\,\,
\bar d'\gamma^\sigma P_{\rm L}^{}d\, \bar\nu\gamma_\sigma^{}P_{\rm L}^{}\nu  \,\,,
\end{eqnarray}
\begin{eqnarray} \label{dd2nn_k}
{\cal H}_{d\bar d'\to\nu\bar\nu}^{\kappa} &=&
\frac{\alpha\, G_{\rm F}^{}}{\sqrt8\, \pi\, \sin^2\theta_{\rm W}^{}} \sum_q
\bigl(\lambda_q^{\rm L}-\lambda_q^{}\bigr)
\biggl( -3\, \ln\frac{\Lambda}{m_W^{}} \,+\, 4 X_0^{}\bigl(x_q^{}\bigr) \biggr)
\bar d'\gamma^\sigma P_{\rm L}^{}d\, \bar\nu\gamma_\sigma^{}P_{\rm L}^{}\nu
\nonumber \\ && \!\!\! +\,\,
\frac{\alpha\, G_{\rm F}^{}}{\sqrt8\, \pi\, \sin^2\theta_{\rm W}^{}} \sum_q \lambda_q^{\rm R}
\biggl[ \bigl(4 x_q^{}-3\bigr) \ln\frac{\Lambda}{m_W^{}} \,+\, \tilde X\bigl(x_q^{}\bigr) \biggr]
\bar d'\gamma^\sigma P_{\rm R}^{}d\, \bar\nu\gamma_\sigma^{}P_{\rm L}^{}\nu \,\,,
\end{eqnarray}
with
\begin{eqnarray} \label{x0}
\begin{array}{c}   \displaystyle
X_0^{}(x) \,\,=\,\, \frac{x(x+2)}{8(x-1)} + \frac{3x(x-2)}{8(x-1)^2}\,\ln x \,\,,
\vspace{2ex} \\   \displaystyle
\tilde X(x) \,\,=\,\, 2 x - \frac{5 x-2 x^2}{1-x}\,\ln x - 4 X_0^{}(x) \,\,.
\end{array}
\end{eqnarray}
In the expressions for ${\cal H}_{d\bar d'\to\nu\bar\nu}$ above, we have neglected the dependence
on the mass of the loop lepton in the box diagram, but it is possible to generalize the formulas
to include the dependence on tau-lepton mass~\cite{paper2}.

The amplitude for \,$d\bar d'\to\ell^+\ell^-$\, gets contributions from all the diagrams
in Fig.~\ref{penguins} via \,$d\bar d'\to\bigl(\gamma^*,Z^*\bigr)\to\ell^+\ell^-$\, and
the diagram in Fig.~\ref{boxes}(a), the loop fermions in the latter being an up-type quark and
a neutrino.  The resulting effective Hamiltonian is
\begin{eqnarray}
{\cal H}_{d\bar d'\to\ell^+\ell^-}^{} \,\,=\,\, {\cal H}_{d\bar d'\to\ell^+\ell^-}^{\rm SM}
\,+\, {\cal H}_{d\bar d'\to\ell^+\ell^-}^{\kappa} \,\,,
\end{eqnarray}
where
\begin{eqnarray} \label{dd2ll_sm}
{\cal H}_{d\bar d'\to\ell^+\ell^-}^{\rm SM} \,\,=\,\,
\frac{\alpha\,G_{\rm F}^{}}{\sqrt8\, \pi} \sum_q 4\lambda_q^{}\, \Biggl(
\frac{-Y_0^{}\bigl(x_q^{}\bigr)}{\sin^2\theta_{\rm W}^{}}\,\,
\bar d'\gamma^\sigma P_{\rm L}^{}d\,\bar\ell\gamma_\sigma^{} P_{\rm L}^{}\ell
\,+\,
2 Z_0^{}\bigl(x_q^{}\bigr)\, \bar d'\gamma^\sigma P_{\rm L}^{}d\, \bar\ell\gamma_\sigma^{}\ell
\Biggr) \,\,,
\end{eqnarray}
\begin{eqnarray} \label{dd2ll_k}
{\cal H}_{d\bar d'\to\ell^+\ell^-}^{\kappa} &=&
\frac{\alpha\, G_{\rm F}^{}}{\sqrt8\, \pi} \sum_q \bigl(\lambda_q^{\rm L}-\lambda_q^{}\bigr) \Biggl[
\biggl( 3\, \ln\frac{\Lambda}{m_W^{}} \,-\, 4 Y_0^{}\bigl(x_q^{}\bigr) \biggr)
\frac{\bar d'\gamma^\sigma P_{\rm L}^{}d\,\bar\ell\gamma_\sigma^{} P_{\rm L}^{}\ell}
     {\sin^2\theta_{\rm W}^{}}
\nonumber \\ && \hspace*{20ex} +\,\,
\biggl(-\frac{16}{3}\, \ln\frac{\Lambda}{m_W^{}} \,+\, 8 Z_0^{}\bigl(x_q^{}\bigr) \biggr)
\bar d'\gamma^\sigma P_{\rm L}^{}d\, \bar\ell\gamma_\sigma^{}\ell \Biggr]
\nonumber \\ && \!\!\! \! +\,\,
\frac{\alpha\, G_{\rm F}^{}}{\sqrt8\, \pi} \sum_q \lambda_q^{\rm R} \Biggl\{
\biggl[ \bigl(3-4 x_q^{}\bigr) \ln\frac{\Lambda}{m_W^{}} \,+\, \tilde Y\bigl(x_q^{}\bigr) \biggr]
\frac{\bar d'\gamma^\sigma P_{\rm R}^{}d\,\bar\ell\gamma_\sigma^{} P_{\rm L}^{}\ell}
     {\sin^2\theta_{\rm W}^{}}
\nonumber \\ && \hspace*{15ex} +\,\,
\biggl[\biggl(8 x_q^{}-\frac{16}{3}\biggr) \ln\frac{\Lambda}{m_W^{}}\,+\,\tilde Z\bigl(x_q^{}\bigr)\biggr]
\bar d'\gamma^\sigma P_{\rm R}^{}d\, \bar\ell\gamma_\sigma^{}\ell \Biggr\} \,\,,
\end{eqnarray}
with  \begin{eqnarray} \label{y0z0}
\begin{array}{c}   \displaystyle
Y_0^{}(x) \,\,=\,\, \frac{x(x-4)}{8(x-1)} + \frac{3x^2}{8(x-1)^2}\,\ln x \,\,,
\vspace{2ex} \\   \displaystyle
Z_0^{}(x) \,\,=\,\, \frac{18x^4-163x^3+259x^2-108x}{144(x-1)^3} +
\frac{24x^4-6x^3-63x^2+50x-8}{72(x-1)^2}\,\ln x \,\,,
\end{array}
\end{eqnarray}
\begin{eqnarray}
\begin{array}{c}   \displaystyle
\tilde Y(x) \,\,=\,\, - 2 x + \frac{5 x-2 x^2}{1-x}\,\ln x + 4 Y_0^{}(x) \,\,, \hspace{5ex}
\tilde Z(x) \,\,=\,\, 2x - 4 x\,\ln x + 8 Z_0^{}(x) \,\,.
\end{array}
\end{eqnarray}
In ${\cal H}_{d\bar d'\to\ell^+\ell^-}^{{\rm SM},\kappa}$ above, we have not displayed terms
contributed by the magnetic $(\sigma^{\mu\nu})$ parts of the  \,$d\bar d'\to\gamma^*,Z^*$\,
amplitudes for convenience, but they can be found in Ref.~\cite{paper2} and do not contribute
to the decay \,$K_L\to\mu^+\mu^-$\, which we consider.
We note that the SM results in Eqs.~(\ref{dd2nn_sm}) and~(\ref{dd2ll_sm})
are in agreement with those found in the literature~\cite{Inami:1980fz,Buchalla:1995vs}.

From the two box-diagrams in Fig.~\ref{boxes} with quarks $d$ and $d'$  in the
external legs and quarks $q$ and $q'$ in the loops, we derive the effective Hamiltonian
\begin{eqnarray}
{\cal H}_{d\bar d'\to\bar d d'}^{} \,\,=\,\, {\cal H}_{d\bar d'\to\bar d d'}^{\rm SM}
\,+\, {\cal H}_{d\bar d'\to\bar d d'}^{\kappa} \,\,,
\end{eqnarray}
where
\begin{eqnarray} \label{dd2dd_sm}
{\cal H}_{d\bar d'\to\bar d d'}^{\rm SM} \,\,=\,\,
\frac{G_{\rm F}^2\,m_W^2}{4\pi^2} \Bigl( \lambda_c^2\,S_0^{}\bigl(x_c^{}\bigr) +
\lambda_t^2\,S_0^{}\bigl(x_t^{}\bigr) + 2 \lambda_c^{}\lambda_t^{}\,S_0^{}\bigl(x_c^{},x_t^{}\bigr)
\Bigr)\, \bar d'\gamma^\alpha P_{\rm L}^{}d\, \bar d'\gamma_\alpha^{}P_{\rm L}^{}d \,\,,
\end{eqnarray}
\begin{eqnarray} \label{dd2dd_k}
{\cal H}_{d\bar d'\to\bar d d'}^\kappa &=&
\frac{G_{\rm F}^2\,m_W^2}{16\pi^2} \sum_{q,q'}
\bigl(\lambda_q^{\rm L}\lambda_{q'}^{\rm L}-\lambda_q^{}\lambda_{q'}^{}\bigr) \Biggl[
\bigl(6-2 x_q^{}\bigr) \ln\frac{\Lambda^2}{m_W^2} \,-\, {\cal B}_1^{}\bigl(x_q^{},x_{q'}^{}\bigr)
\Biggr]\, \bar d'\gamma^\alpha P_{\rm L}^{}d\,\bar d'\gamma_\alpha^{}P_{\rm L}^{}d
\nonumber \\ && \!\!\!\! +\,\,
\frac{G_{\rm F}^2\,m_W^2}{4\pi^2} \sum_{q,q'} \lambda_q^{}\lambda_{q'}^{\rm R}\, \biggl[
\bigl(6-x_q^{}-x_{q'}^{}\bigr) \ln\frac{\Lambda^2}{m_W^2} \,-\, {\cal B}_2^{}\bigl(x_q^{},x_{q'}^{}\bigr)
\biggr]\, \bar d'\gamma^\alpha P_{\rm L}^{}d\,\bar d'\gamma_\alpha^{}P_{\rm R}^{}d
\nonumber \\ && \!\!\!\! +\,\,
\frac{G_{\rm F}^2\,m_W^2}{4\pi^2} \sum_{q,q'} \lambda_q^{}\lambda_{q'}^{}\sqrt{x_q^{}x_{q'}^{}}
\biggl( -\ln\frac{\Lambda^2}{m_W^2} \,-\, {\cal B}_3^{}\bigl(x_q^{},x_{q'}^{}\bigr) \biggr)
\nonumber \\ && \times \,\,
\Bigl( \kappa_{qd}^{\rm R}\kappa_{q'd}^{\rm R}\, \bar d'P_{\rm R}^{}d\,\bar d'P_{\rm R}^{}d +
      \kappa_{qd'}^{\rm R*}\kappa_{q'd'}^{\rm R*}\, \bar d'P_{\rm L}^{}d\,\bar d'P_{\rm L}^{}d \Bigr) \,\,,
\end{eqnarray}
with
\begin{eqnarray}
S_0^{}(x,y) \,\,=\,\, \frac{-3x y}{4(1-x)(1-y)} -
\frac{x y\bigl(4-8 x+x^2\bigr) \ln x}{4(y-x)(1-x)^2} -
\frac{x y\bigl(4-8 y+y^2\bigr) \ln y}{4(x-y)(1-y)^2} \,\,,
\end{eqnarray}
\begin{eqnarray} \label{123}
\begin{array}{c}   \displaystyle
{\cal B}_1^{}(x,y) \,\,=\,\, \frac{3}{2}(x+y) + \frac{3(x+y-x y)}{(1-x)(1-y)}
+ \frac{\bigl(4 x^2-8 x^3+x^4\bigr) \ln x}{(y-x)(1-x)^2}
+ \frac{\bigl(4 y^2-8 y^3+y^4\bigr) \ln y}{(x-y)(1-y)^2} \,\,,
\vspace{2ex} \\   \displaystyle
{\cal B}_2^{}(x,y) \,\,=\,\, \frac{3}{2}(x+y) - \frac{9(x+y-x y)}{(1-x)(1-y)} +
\frac{(4-x)^2 x^2\, \ln x}{(y-x)(1-x)^2} + \frac{(4-y)^2 y^2\, \ln y}{(x-y)(1-y)^2} \,\,,
\vspace{2ex} \\   \displaystyle
{\cal B}_3^{}(x,y) \,\,=\,\,
\frac{x y-x-y-2}{(1-x)(1-y)} + \frac{\bigl(4 x-2 x^2+x^3) \ln x}{(y-x)(1-x)^2} +
\frac{\bigl(4 y-2 y^2+y^3\bigr) \ln y}{(x-y)(1-y)^2} \,\,,
\end{array}
\end{eqnarray}
and  \,$S_0^{}(x)=\lim_{y\to x}S_0^{}(x,y)$.\,
The expression for ${\cal H}_{d\bar d'\to\bar d d'}^{\rm SM}$ agrees with that in
the literature~\cite{Inami:1980fz,Buchalla:1995vs}.

\section{Matrix elements and parameters\label{me}}

The matrix elements used in estimating the amplitudes for  $\bar B\to J/\psi\bar K,\eta_c^{}\bar K$ are
\begin{eqnarray}
\begin{array}{c}   \displaystyle
\bigl\langle\psi\bigr|\bar c_m^{}\gamma^\mu c_n^{}\bigl|0\bigr\rangle
\,\,=\,\,  \frac{\delta_{mn}^{}}{N_{\rm c}^{}}\, f_\psi^{} m_\psi^{}\, \varepsilon^\mu \,\,,
\hspace{2em}
\bigl\langle\bar K^0(q)\bigr|\bar s_m^{}b_n^{}\bigl|\bar B^0(p)\bigr\rangle \,\,=\,\,
\frac{\delta_{mn}^{}}{N_{\rm c}^{}}\, \frac{p^2-q^2}{m_b^{}-m_s^{}}\, F_0^{BK} \;,
\vspace{1ex} \\   \displaystyle
\bigl\langle\bar K^0(q)\bigr|\bar s_m^{}\gamma_\mu^{}b_n^{}\bigl|\bar B^0(p)\bigr\rangle
\,\,=\,\,
\frac{\delta_{mn}^{}}{N_{\rm c}^{}} \biggl[ (p+q)^\mu\,  F_1^{BK} +
\frac{(p-q)^\mu}{(p-q)^2}\bigl(m_B^2-m_K^2\bigr)\bigl(F_0^{BK}-F_1^{BK}\bigr) \biggr] \,\,,
\vspace{1ex} \\   \displaystyle
\bigl\langle\eta_c^{}\bigr|\bar c_m^{}\gamma^\mu\gamma_5^{}c_n^{}\bigl|0\bigr\rangle \,\,=\,\,
\frac{\delta_{mn}^{}}{N_{\rm c}^{}}\, i f_{\eta_c^{}}^{}\, p_{\eta_c^{}}^\mu \,\,, \hspace{2em}
\bigl\langle\eta_c^{}\bigr|\bar c_m^{}\gamma_5^{}c_n^{}\bigl|0\bigr\rangle \,\,=\,\,
\frac{\delta_{mn}^{}}{N_{\rm c}^{}}\, \frac{i f_{\eta_c^{}}^{}\,m_{\eta_c^{}}^2}{2 m_c^{}} \,\,,
\end{array}
\end{eqnarray}
where  \,$N_{\rm c}^{}=3$\, is the number of colors  and  $m$ and $n$ are color indices.
The decay constants above are  \,$f_\psi^{}=416$\,MeV\, extracted from \,$\Gamma(J/\psi\to e^+e^-)$\,
data~\cite{pdg}  and  \,$f_{\eta_c^{}}=420$\,MeV\, calculated in Ref.~\cite{Hwang:1997ie}.
The form factors  $F_{0,1}^{BK}$ are functions of $(p-q)^2$, with
\,$F_0^{BK}\bigl(m_{\eta_c}^2\bigr)=0.45$\,  and
\,$F_1^{BK}\bigl(m_\psi^2\bigr)=0.65$\,  from Ref.~\cite{Cheng:2003sm}.
For meson masses, we use the values in Ref.~\cite{pdg}.

The $K$, $B_d$, and $B_s$ decay-constants are~\cite{pdg,ckmfit}
\begin{eqnarray}
f_K^{} \,\,=\,\, 155.5\pm 0.8 \,\,, \hspace{5ex}
f_{B_d} \,\,=\,\, 191\pm 15 \,\,, \hspace{5ex}
f_{B_s} \,\,=\,\, 228\pm 17 \,\,,
\end{eqnarray}
all in units of MeV.
All of the following parameter values are obtained from Ref.~\cite{ckmfit}, the experimental
and theoretical errors given therein having been combined in quadrature.
The QCD-correction factors in the $K$- and $B_{d,s}$-mixing amplitudes
are~\cite{Buchalla:1995vs,ckmfit,Herrlich:1993yv}.
\begin{eqnarray}
\eta_{cc}^{} \,\,=\,\, 1.46\pm 0.22 \,\,, \hspace{5ex}
\eta_B^{} \,\,=\,\, 0.551\pm0.007 \,\,.
\end{eqnarray}
The bag parameters used in the $K$-mixing amplitude are defined by
\begin{eqnarray}
\bigl\langle K^0\bigr|\bar d\gamma^\alpha P_{\rm L}^{}s\,
\bar d\gamma_\alpha^{}P_{\rm L}s\bigl|\bar K^0\bigr\rangle \,\,=\,\,
\frac{2}{3}\,f_K^2 m_K^2\, B_K^{} \,\,, \hspace{4ex}
\bigl\langle K^0\bigr|\bar d\gamma^\alpha P_{\rm L}^{}s\,
\bar d\gamma_\alpha^{}P_{\rm R}s\bigl|\bar K^0\bigr\rangle \,\,=\,\,
\frac{-f_K^2\, m_K^4\, B_K^{}}{3\bigl(m_d^{}+m_s^{}\bigr)^2} \,\,, \;\;\;
\end{eqnarray}
and similarly for $B_{B_{d,s}}$ in the $B_{d,s}^{}$-mixing cases, where~\cite{ckmfit}
\begin{eqnarray}
B_K^{} \,\,=\,\, 0.72\pm 0.04 \,\,, \hspace{5ex}
B_{B_d}^{} \,\,=\,\, 1.17\pm 0.08 \,\,, \hspace{5ex}
B_{B_s}^{} \,\,=\,\, 1.23\pm 0.06\,\,.
\end{eqnarray}
The charm and top masses used in the loop functions are~\cite{ckmfit,Allison:2008xk}
\begin{eqnarray}
\bar m_c^{}\bigl(m_c^{}\bigr) \,\,=\,\, 1.29\pm 0.04 \,\,, \hspace{5ex}
\bar m_t^{}\bigl(m_t^{}\bigr) \,\,=\,\, 165\pm1 \,\,,
\end{eqnarray}
both in units of GeV.
For the CKM parameters, we adopt in the Wolfenstein parametrization the central values~\cite{ckmfit}
\begin{eqnarray}
A \,\,=\,\, 0.8116 \,\,, \hspace{5ex} \lambda \,\,=\,\, 0.22521 \,\,, \hspace{5ex}
\bar\rho \,\,=\,\, 0.139 \,\,, \hspace{5ex} \bar\eta \,\,=\,\, 0.341 \,\,.
\end{eqnarray}
In our numerical estimates, we use only the central values of the parameters above, as
their errors being no more than 20\% are within the intrinsic uncertainty of our analysis.


\begin{thebibliography}{99}

\bibitem{Buchmuller:1985jz}
  W.~Buchmuller and D.~Wyler,
  Nucl.\ Phys.\  B {\bf 268}, 621 (1986).

\bibitem{Peccei:1989kr}
  R.D.~Peccei and X.~Zhang,
  Nucl.\ Phys.\  B {\bf 337}, 269 (1990);
  R.D.~Peccei, S.~Peris, and X.~Zhang,
  Nucl.\ Phys.\  B {\bf 349}, 305 (1991).

\bibitem{Fujikawa:1993zu}
  K.~Fujikawa and A.~Yamada,
  Phys.\ Rev.\  D {\bf 49}, 5890 (1994);
  F.~Larios, M.A.~Perez, and C.P.~Yuan,
  Phys.\ Lett.\  B {\bf 457}, 334 (1999)
  [arXiv:hep-ph/9903394];
  G.~Burdman, M.C.~Gonzalez-Garcia, and S.F.~Novaes,
  Phys.\ Rev.\  D {\bf 61}, 114016 (2000)
  [arXiv:hep-ph/9906329];
  B.~Grzadkowski and M.~Misiak,
  Phys.\ Rev.\  D {\bf 78}, 077501 (2008)
  [arXiv:0802.1413 [hep-ph]];
  J.P.~Lee and K.Y.~Lee,
  Phys.\ Rev.\  D {\bf 78}, 056004 (2008)
  [arXiv:0806.1389 [hep-ph]].

\bibitem{anomtqc}
  D.O.~Carlson, E.~Malkawi, and C.P.~Yuan,
  Phys.\ Lett.\  B {\bf 337}, 145 (1994)
  [arXiv:hep-ph/9405277];
  E.~Malkawi and C.P.~Yuan,
  Phys.\ Rev.\  D {\bf 50}, 4462 (1994)
  [arXiv:hep-ph/9405322].

\bibitem{Kagan:1994qg}
  A.L.~Kagan,
  Phys.\ Rev.\  D {\bf 51}, 6196 (1995)
  [arXiv:hep-ph/9409215];
  G.~Colangelo, G.~Isidori, and J.~Portoles,
  Phys.\ Lett.\  B {\bf 470}, 134 (1999)
  [arXiv:hep-ph/9908415];
  G.~D'Ambrosio, G.~Isidori, and G.~Martinelli,
  Phys.\ Lett.\  B {\bf 480}, 164 (2000)
  [arXiv:hep-ph/9911522];
  J.~Tandean and G.~Valencia,
  Phys.\ Rev.\  D {\bf 62}, 116007 (2000)
  [arXiv:hep-ph/0008238];
  D.N.~Gao,
  Phys.\ Rev.\  D {\bf 67}, 074028 (2003)
  [arXiv:hep-ph/0212280];
  F.~Mescia, C.~Smith, and S.~Trine,
  JHEP {\bf 0608}, 088 (2006)
  [arXiv:hep-ph/0606081].

\bibitem{Buras:1999da}
  A.J.~Buras, G.~Colangelo, G.~Isidori, A.~Romanino, and L.~Silvestrini,
  Nucl.\ Phys.\  B {\bf 566}, 3 (2000)
  [arXiv:hep-ph/9908371].

\bibitem{Ecker:1983dj}
  G.~Ecker, W.~Grimus, and H.~Neufeld,
  Nucl.\ Phys.\  B {\bf 229}, 421 (1983).

\bibitem{He:1989xj}
  X.G.~He, B.H.J.~McKellar, and S.~Pakvasa,
  Int.\ J.\ Mod.\ Phys.\  A {\bf 4}, 5011 (1989)
  [Erratum-ibid.\  A {\bf 6}, 1063 (1991)];
  Phys.\ Lett.\  B {\bf 254}, 231 (1991).

\bibitem{Ginges:2003qt}
  J.S.M.~Ginges and V.V.~Flambaum,
  Phys.\ Rept.\  {\bf 397}, 63 (2004)
  [arXiv:physics/0309054].

\bibitem{Dib:2006hk}
  C.~Dib, A.~Faessler, T.~Gutsche, S.~Kovalenko, J.~Kuckei, V.E.~Lyubovitskij, and K.~Pumsa-ard,
  J.\ Phys.\ G {\bf 32}, 547 (2006)
  [arXiv:hep-ph/0601144]; references therein.

\bibitem{Georgi:1994qn}
  H.~Georgi,
  Ann.\ Rev.\ Nucl.\ Part.\ Sci.\  {\bf 43} (1993) 209.

\bibitem{He:2004it}
  X.G.~He and G.~Valencia,
  Phys.\ Rev.\  D {\bf 70}, 053003 (2004)
  [arXiv:hep-ph/0404229];
  X.G.~He, G.~Valencia, and Y.~Wang,
  Phys.\ Rev.\  D {\bf 70}, 113011 (2004)
  [arXiv:hep-ph/0409346].

\bibitem{pdg}
  C.~Amsler {\it et al.}  [Particle Data Group],
  Phys.\ Lett.\  B {\bf 667}, 1 (2008).

\bibitem{Stone:2008gw}
  S.~Stone,
  arXiv:0806.3921 [hep-ex];
  J.P.~Alexander  [CLEO Collaboration],
  arXiv:0901.1216 [hep-ex].

\bibitem{Dobrescu:2008er}
  B.A.~Dobrescu and A.S.~Kronfeld,
  Phys.\ Rev.\ Lett.\  {\bf 100}, 241802 (2008)
  [arXiv:0803.0512 [hep-ph]];
  A.S.~Kronfeld,
  PoS LATTICE2008, 282 (2008)
  [arXiv:0812.2030 [hep-lat]].

\bibitem{Narison:2008bc}
  S.~Narison,
  Phys.\ Lett.\  B {\bf 668}, 308 (2008)
  [arXiv:0807.2830 [hep-ph]].

\bibitem{Kowalewski:2008zz}
  R.~Kowalewski and T.~Mannel, in Ref.~\cite{pdg}.

\bibitem{Isgur:1989vq}
  N.~Isgur and M.B.~Wise,
  Phys.\ Lett.\  B {\bf 232}, 113 (1989);
A.V.~Manohar and M.B.~Wise, {\it Heavy Quark Physics}
(Cambridge University Press, Cambridge, 2000).

\bibitem{hfag}
Heavy Flavor Averaging Group, http://www.slac.stanford.edu/xorg/hfag.

\bibitem{Nir:1992wi}
  Y.~Nir and H.R.~Quinn,
  Ann.\ Rev.\ Nucl.\ Part.\ Sci.\  {\bf 42}, 211 (1992).

\bibitem{Buchalla:1995vs}
  G.~Buchalla, A.J.~Buras, and M.E.~Lautenbacher,
  Rev.\ Mod.\ Phys.\  {\bf 68}, 1125 (1996)
  [arXiv:hep-ph/9512380].

\bibitem{Cho:1993zb}
  P.L.~Cho and M.~Misiak,
  Phys.\ Rev.\  D {\bf 49}, 5894 (1994)
  [arXiv:hep-ph/9310332].

\bibitem{nceff}
  M.~Bauer, B.~Stech, and M.~Wirbel,
  Z.\ Phys.\  C {\bf 34}, 103 (1987);
  A.~Ali and C.~Greub,
  Phys.\ Rev.\  D {\bf 57}, 2996 (1998)
  [arXiv:hep-ph/9707251];
  H.Y.~Cheng and B.~Tseng,
  Phys.\ Rev.\  D {\bf 58}, 094005 (1998)
  [arXiv:hep-ph/9803457];
  A.~Ali, G.~Kramer, and C.D.~Lu,
  Phys.\ Rev.\  D {\bf 58}, 094009 (1998)
  [arXiv:hep-ph/9804363].

\bibitem{He:1999ik}
  X.G.~He and G.~Valencia,
  Phys.\ Rev.\  D {\bf 61}, 075003 (2000)
  [arXiv:hep-ph/9908298].

\bibitem{Donoghue:1985ae}
  J.F.~Donoghue and B.R.~Holstein,
  Phys.\ Rev.\  D {\bf 32}, 1152 (1985);
  X.G.~He and G.~Valencia,
  Phys.\ Rev.\  D {\bf 52}, 5257 (1995)
  [arXiv:hep-ph/9508411].

\bibitem{He:1999bv}
  X.G.~He, H.~Murayama, S.~Pakvasa, and G.~Valencia,
  Phys.\ Rev.\  D {\bf 61}, 071701 (2000)
  [arXiv:hep-ph/9909562].

\bibitem{Tandean:2003fr}
  J.~Tandean,
  Phys.\ Rev.\  D {\bf 69}, 076008 (2004)
  [arXiv:hep-ph/0311036].

\bibitem{ckmfit}
CKMfitter, http://ckmfitter.in2p3.fr.

\bibitem{Eeg:2008kf}
  J.O.~Eeg, K.~Kumericki, and I.~Picek,
  Phys.\ Lett.\  B {\bf 669}, 150 (2008)
  [arXiv:0803.2106 [hep-ph]];
  A.J.~Buras and D.~Guadagnoli,
  arXiv:0901.2056 [hep-ph].

\bibitem{Donoghue:1986hh}
  J.F.~Donoghue, X.G.~He, and S.~Pakvasa,
  Phys.\ Rev.\  D {\bf 34}, 833 (1986);
  X.G.~He, H.~Steger, and G.~Valencia,
  Phys.\ Lett.\  B {\bf 272}, 411 (1991);
N.G.~Deshpande, X.G.~He, and S.~Pakvasa,
{\it ibid.}~{\bf 326}, 307 (1994);
  J.~Tandean and G.~Valencia,
  Phys.\ Rev.\  D {\bf 67}, 056001 (2003)
  [arXiv:hep-ph/0211165].

\bibitem{hypercp}
C. Materniak [HyperCP Collaboration],
Talk given at the Eighth International Conference on Hyperons, Charm and Beauty Hadrons,
22-28 June 2008,  Columbia, South Carolina.

\bibitem{Pondrom:1981gu}
  L.~Pondrom {\it et al.},
  Phys.\ Rev.\  D {\bf 23}, 814 (1981).

\bibitem{Inami:1980fz}
  T.~Inami and C.~S.~Lim,
  Prog.\ Theor.\ Phys.\  {\bf 65}, 297 (1981)
  [Erratum-ibid.\  {\bf 65}, 1772 (1981)].

\bibitem{Buras:2006gb}
  A.J.~Buras, M.~Gorbahn, U.~Haisch, and U.~Nierste,
  JHEP {\bf 0611}, 002 (2006)
  [arXiv:hep-ph/0603079];
  F.~Mescia and C.~Smith,
  Phys.\ Rev.\  D {\bf 76}, 034017 (2007)
  [arXiv:0705.2025 [hep-ph]];
  A.J.~Buras, F.~Schwab, and S.~Uhlig,
  Rev.\ Mod.\ Phys.\  {\bf 80}, 965 (2008)
  [arXiv:hep-ph/0405132];
  J.~Brod and M.~Gorbahn,
  Phys.\ Rev.\  D {\bf 78}, 034006 (2008)
  [arXiv:0805.4119 [hep-ph]].

\bibitem{Artamonov:2008qb}
  A.V.~Artamonov {\it et al.}  [E949 Collaboration],
  Phys.\ Rev.\ Lett.\  {\bf 101}, 191802 (2008)
  [arXiv:0808.2459 [hep-ex]].

\bibitem{Gorbahn:2006bm}
  M.~Gorbahn and U.~Haisch,
  Phys.\ Rev.\ Lett.\  {\bf 97}, 122002 (2006)   [arXiv:hep-ph/0605203].

\bibitem{Littenberg:2008zz}
  L.~Littenberg and G.~Valencia, in Ref.~\cite{pdg}.

\bibitem{AbdElHady:1997eu}
  A.~Abd El-Hady and G.~Valencia,
  Phys.\ Lett.\  B {\bf 414}, 173 (1997)
  [arXiv:hep-ph/9704300].

\bibitem{Chang:1994wk}
  D.~Chang, X.G.~He, and S.~Pakvasa,
  Phys.\ Rev.\ Lett.\  {\bf 74}, 3927 (1995)
  [arXiv:hep-ph/9412254];
  A.~Masiero and H.~Murayama,
  Phys.\ Rev.\ Lett.\  {\bf 83}, 907 (1999)
  [arXiv:hep-ph/9903363].

\bibitem{Langacker:1989xa}
  P.~Langacker and S.~Uma Sankar,
  Phys.\ Rev.\  D {\bf 40}, 1569 (1989);
  X.G.~He and G.~Valencia,
  Phys.\ Rev.\  D {\bf 66}, 013004 (2002) [Erratum-ibid.\  D {\bf 66}, 079901 (2002)]
  [arXiv:hep-ph/0203036].

\bibitem{paper2}
X.G.~He, J.~Tandean, and G.~Valencia, in preparation.

\bibitem{Hwang:1997ie}
  D.S.~Hwang and G.H.~Kim,
  Z.\ Phys.\  C {\bf 76}, 107 (1997)
  [arXiv:hep-ph/9703364].

\bibitem{Cheng:2003sm}
H.Y.~Cheng, C.K.~Chua, and C.W.~Hwang,
Phys.\ Rev.\  D {\bf 69}, 074025 (2004)
[arXiv:hep-ph/0310359].

\bibitem{Herrlich:1993yv}
  S.~Herrlich and U.~Nierste,
  Nucl.\ Phys.\  B {\bf 419}, 292 (1994)
  [arXiv:hep-ph/9310311];
  J.H.~Kuhn, M.~Steinhauser, and C.~Sturm,
  Nucl.\ Phys.\  B {\bf 778}, 192 (2007)
  [arXiv:hep-ph/0702103];

\bibitem{Allison:2008xk}
  I.~Allison {\it et al.}  [HPQCD Collaboration],
  Phys.\ Rev.\  D {\bf 78}, 054513 (2008)
  [arXiv:0805.2999 [hep-lat]];
  M.~Steinhauser,
  arXiv:0809.1925 [hep-ph].

\end{thebibliography}
\end{document}